\documentclass[useAMS,usenatbib,times,letter,amssymb]{mn2e}
\usepackage{epsfig,times,amssymb,amsmath,color,verbatim}

\title[CFHTLenS: Tomographic weak lensing]{CFHTLenS tomographic weak lensing cosmological parameter constraints:  Mitigating the impact of intrinsic galaxy alignments}
\author[C. Heymans et al.]{Catherine Heymans$^{1}$\thanks{heymans@roe.ac.uk},  
Emma Grocutt$^1$, Alan Heavens$^{2,1}$, Martin Kilbinger$^{3,4,5,6}$,
\newauthor Thomas D. Kitching$^{7,1}$, Fergus Simpson$^1$, 
Jonathan Benjamin$^8$, Thomas Erben$^9$, 
\newauthor Hendrik Hildebrandt$^{8,9}$, Henk Hoekstra$^{10,11}$, 
Yannick Mellier$^{6,3}$, Lance Miller$^{12}$, 
\newauthor Ludovic Van Waerbeke$^8$, 
Michael L. Brown$^{13}$, Jean Coupon$^{14}$, Liping Fu$^{15}$, 
\newauthor Joachim Harnois-D\'{e}raps$^{16,17}$, 
Michael J. Hudson$^{18,19}$, 
Konrad Kuijken$^{10}$, 
\newauthor Barnaby Rowe$^{20,21}$, 
Tim Schrabback$^{9,22,10}$,  
Elisabetta Semboloni$^{10}$, 
Sanaz Vafaei$^{8}$,
\newauthor Malin Velander$^{12,10}.$\\
\\
$^1$Scottish Universities Physics Alliance, Institute for Astronomy, University of Edinburgh, Royal Observatory, Blackford Hill, Edinburgh, EH9 3HJ, UK.\\ 
$^2$Imperial Centre for Inference and Cosmology, Imperial College London, Blackett Laboratory, Prince Consort Road, London, SW7 2AZ, UK.\\ 
$^{3}$CEA Saclay, Service d'Astrophysique (SAp), Orme des Merisiers, B\^at 709, F-91191 Gif-sur-Yvette, France.\\
$^{4}$Excellence Cluster Universe, Boltzmannstr. 2, D-85748 Garching, Germany.\\
$^{5}$Universit\"ats-Sternwarte, Ludwig-Maximillians-Universit\"at M\"unchen, Scheinerstr.~1, 81679 M\"unchen, Germany.\\
$^6$Institut d'Astrophysique de Paris, UniversitŽ Pierre et Marie Curie - Paris 6, 98 bis Boulevard Arago, F-75014 Paris, France.\\
$^7$Mullard Space Science Laboratory, University College London, Holmbury St Mary, Dorking, Surrey RH5 6NT, UK.\\
$^8$Department of Physics and Astronomy, University of British Columbia, 6224 Agricultural Road, Vancouver, V6T 1Z1, BC, Canada.\\  
$^9$Argelander Institute for Astronomy, University of Bonn, Auf dem H{\"u}gel 71, 53121 Bonn, Germany.\\
$^{10}$Leiden Observatory, Leiden University, Niels Bohrweg 2, 2333 CA Leiden, The Netherlands.\\
$^{11}$Department of Physics and Astronomy, University of Victoria, Victoria, BC V8P 5C2, Canada.\\
$^{12}$Department of Physics, Oxford University, Keble Road, Oxford OX1 3RH, UK.\\ 
$^{13}$Jodrell Bank Centre for Astrophysics, University of Manchester, Oxford Road, Manchester, M13 9PL, UK. \\ 
$^{14}$Institute of Astronomy and Astrophysics, Academia Sinica, P.O. Box 23-141, Taipei 10617, Taiwan.\\
$^{15}$Key Lab for Astrophysics, Shanghai Normal University, 100 Guilin Road, 200234, Shanghai, China. \\
$^{16}$Canadian Institute for Theoretical Astrophysics, University of Toronto, M5S 3H8, Ontario, Canada.\\
$^{17}$Department of Physics, University of Toronto, M5S 1A7, Ontario, Canada.\\
$^{18}$Department of Physics and Astronomy, University of Waterloo, Waterloo, ON, N2L 3G1, Canada.\\
$^{19}$Perimeter Institute for Theoretical Physics, 31 Caroline Street N, Waterloo, ON, N2L 1Y5, Canada.\\
$^{20}$Department of Physics and Astronomy, University College London, Gower Street, London WC1E 6BT, UK.\\
$^{21}$California Institute of Technology, 1200 E California Boulevard, Pasadena CA 91125, USA.\\
$^{22}$Kavli Institute for Particle Astrophysics and Cosmology, Stanford University, 382 Via Pueblo Mall, Stanford, CA 94305-4060, USA.\\
}
\newcommand{\be}{\begin{equation}}  \newcommand{\ee}{\end{equation}}
  
\newcommand{\ba}{\begin{eqnarray}}\newcommand{\ea}{\end{eqnarray}}
\newcommand{\bm}[1]{\mbox{\boldmath{$#1$}}}   

\def\gs{\mathrel{\raise1.16pt\hbox{$>$}\kern-7.0pt %
\lower3.06pt\hbox{{$\scriptstyle \sim$}}}}         %
\def\ls{\mathrel{\raise1.16pt\hbox{$<$}\kern-7.0pt %
\lower3.06pt\hbox{{$\scriptstyle \sim$}}}}         %

\begin{document}

\maketitle

\begin{abstract}

We present a finely-binned tomographic weak lensing analysis of the Canada-France-Hawaii Telescope Lensing Survey, CFHTLenS,
mitigating contamination to the signal from the presence of intrinsic galaxy alignments via the simultaneous fit of a cosmological model and an intrinsic alignment model.  CFHTLenS spans 154 square degrees in five optical bands, with accurate shear and photometric redshifts for a galaxy sample with a median redshift of $z_{\rm m} =0.70$. 
We estimate the 21 sets of cosmic shear correlation functions associated with six redshift bins, each spanning the angular range of  $1.5<\theta<35$ arcmin. We combine this CFHTLenS data with auxiliary cosmological probes: the cosmic microwave background with data from WMAP7, baryon acoustic oscillations with data from BOSS, and a prior on the Hubble constant from the HST distance ladder.
This leads to constraints on the normalisation of the matter power spectrum $\sigma_8 = 0.799 \pm 0.015$ and the matter density parameter $\Omega_{\rm m} = 0.271 \pm 0.010$ for a flat $\Lambda$CDM cosmology.  For a flat $w$CDM cosmology we constrain the dark energy equation of state parameter $w = -1.02 \pm 0.09$.   We also provide constraints for curved $\Lambda$CDM  and $w$CDM cosmologies.
We find the intrinsic alignment contamination to be galaxy-type dependent with a significant intrinsic alignment signal found for early-type galaxies, in contrast to the late-type galaxy sample for which the intrinsic alignment signal is found to be consistent with zero.

\end{abstract}


\section{Introduction}
\label{sec:intro}
Cosmological weak lensing is the study of the weak gravitational distortions imprinted on the images of distant galaxies by large-scale structures.  A series of deflections are induced by the gravitational potential of the dark and luminous matter that light passes, as it travels through the Universe.  This lensing effect results in a coherent distortion, detected in the observed images of galaxies, that allows us to infer the distribution and density of matter in the Universe.  This well understood physical effect is recognized as one of the most powerful probes of cosmology, allowing not only the direct study of dark matter, but also, through the study of the growth of structures, a unique probe of gravity and dark energy on large scales \citep[see review by][and references therein for more details]{Weinberg}.

It has long been recognized that the optimal way to extract cosmological information from the detection of weak gravitational lensing is to utilize redshift information, for example by separating the lensed galaxy sample into a number of tomographic bins using photometric redshift information \citep{Hu99}.  This idea was explored theoretically to determine the optimal number of redshift bins \citep[see for example][]{Huterer02,SKS04} and also significantly extended to consider a fully three-dimensional analysis \citep{AFH03}.  These early predictions suggested that it was possible to achieve up to an order-of-magnitude improvement on cosmological constraints when weak lensing measurements were made in combination with photometric redshifts.  This drove weak lensing survey designs to include multi-band optical imaging for redshift estimates, in addition to high-resolution imaging in a single band for the measurement of the weak lensing distortions.

In order to detect cosmological weak lensing, correlations are measured between the shapes of galaxies whose observed ellipticity, $\epsilon_{\rm obs}$, is related to their intrinsic ellipticity, $\epsilon_{\rm s}$, and the weak cosmological shear distortion that we wish to extract, $\gamma$, through\footnote{For clarity we assume this simplified form to relate galaxy ellipticity and shear.  In detail, the relationship also depends on the ellipticity estimator used in the analysis and how rapidly the galaxy ellipticity varies at different isophotal limits (see Section~\ref{sec:theory}).  We also assume here the weak shear limit that $|\gamma| \ll 1$.}
\be
\epsilon_{\rm obs} = \epsilon_{\rm s} + \gamma \, .
\label{eqn:egamma}
\ee
The observed angular two-point correlation function $\langle \epsilon_{\rm obs}^a \epsilon_{\rm obs}^b \rangle$ is then determined by averaging over all galaxy pairs $(a,b)$ separated by angle $\theta$.  This observed quantity is related to the angular two-point shear correlation function $\langle \gamma^a \gamma^b \rangle$ through
\be
\langle \epsilon_{\rm obs}^a \epsilon_{\rm obs}^b \rangle  = 
\langle \epsilon_{\rm s}^a \epsilon_{\rm s}^b \rangle +
\langle \epsilon_{\rm s}^a \gamma^b \rangle +
\langle \gamma^a \epsilon_{\rm s}^b \rangle +
\langle \gamma^a \gamma^b \rangle  \, .
\label{eqn:eeIIGIGG}
\ee
For the majority of weak lensing analyses to date, the first three terms on the right hand side of equation~\ref{eqn:eeIIGIGG} have been assumed to be sufficiently small to be ignored (although see discussion below for analyses that do not make this assumption).  In this case the observed angular two-point correlation function $\langle \epsilon_{\rm obs}^a \epsilon_{\rm obs}^b \rangle$ is equated with the cosmological shear correlation function $\langle \gamma^a \gamma^b \rangle$, which can be directly related to the underlying matter power spectrum of density fluctuations in the Universe (see Section~\ref{sec:method} for more details).    It is, however, possible that these intrinsic terms are significantly non-zero, arising from correlations induced during galaxy formation,  between a galaxy's intrinsic shape and its local density field.  It was noted early on that such an effect would be significantly more detrimental to the future measurement of weak lensing in tomographic bins \citep{KingSch03, HBH04}, in comparison to the standard two-dimensional analysis of that time.

The study of the impact of intrinsic galaxy alignments on weak lensing studies initially focussed only on the intrinsic alignment of physically nearby galaxies; $\langle \epsilon_{\rm s}^a \epsilon_{\rm s}^b \rangle$, hereafter referred to as `II'.  The broad agreement in results between the first estimates from numerical simulations \citep{HRH00,CM00}, analytical studies \citep{CKB01,CNPT01,LP01}, and the first low redshift observational constraints \citep{PLS00,BTHD02} resulted in a consistent picture; for deep weak lensing surveys, the contamination to the weak lensing signal was expected to be less than a few per cent effect.  \citet{HS04} were the first to highlight, however, the importance of also including the shear-shape correlations in the analysis, $\langle \epsilon_{\rm s} \gamma \rangle$, for galaxies that are separated by large physical distances along the line of sight.   In this case the background galaxy experiences a shear $\gamma$ caused by the foreground tidal gravitational field.  If the foreground galaxy has an intrinsic ellipticity that is linearly correlated with this field, $\langle \epsilon_{\rm s} \gamma \rangle$ is no longer expected to be zero.  We refer to this effect hereafter as `GI'.  

The most recent observational results have focused on inferring the amplitude of the II and GI signal by measuring the local cross-correlation between galaxy number densities and ellipticities, to determine the correlation between galaxy shape and the local density field.  This method was first implemented using low redshift $ z \sim 0.1$ galaxies in the Sloan Digital Sky Survey spectroscopic sample \citep[SDSS,][]{RM06}.  It was then extended to higher redshifts using the SDSS Luminous Red Galaxy sample (LRG) between $0.15<z<0.35$ \citep{Hirata07}, and the MegaZ-LRG sample out to $z \sim 0.6$ \citep{BJ11}.  In these extensive studies, a clear dependence on galaxy type is detected, with the most massive red galaxies exhibiting the strongest intrinsic correlations.  \citet{Hirata07} show their data match models for this galaxy-type dependence as predicted from numerical simulations \citep{HeymansIA06}.  The massive LRG population is, however, in the minority when it comes to a typical deep cosmological weak lensing survey, which is dominated by the blue galaxy population.  The most representative observational analysis of this effect therefore comes from the highest redshift measurements of the GI and II effect, using blue galaxies from the WiggleZ survey out to $z \sim 0.7$ \citep{RM11}.  The reported null detection of this effect for blue galaxies results in a predicted systematic error of at most $\pm 0.03$ on the amplitude of the matter power spectrum, $\sigma_8$, for a non-tomographic analysis of a survey like the Canada-France-Hawaii Telescope Lensing Survey (CFHTLenS).  We perform a tomographic weak lensing analysis of this survey in this paper.    

For a deep survey, such as CFHTLenS, the correlation between a galaxy's shape and its local density field impacts more strongly on a tomographic analysis than a two-dimensional analysis for the following reasons.  For a deep two-dimensional analysis, there is only a small fraction of galaxy pairs, at fixed angular separation $\theta$, that are physically close enough to have formed in the same density field and hence experience some degree of alignment.  The significant II signal from these close pairs is therefore greatly diluted as the majority of galaxy pairs are well separated in three-dimensions.  The lensing of background galaxies by a foreground structure is most efficient when the foreground structure is well separated from the source.  In a two-dimensional analysis, the signal from the galaxy pairs that experience the strongest GI effects is therefore again diluted by the presence of closer galaxy pairs in the analysis where the lensing effect is less efficient.  In contrast, a tomographic analysis will include redshift bin combinations that, in fact, enhance the contamination by both these effects.  The auto-correlation measurement of the lensing signal within narrow redshift bins dramatically increases the fraction of physically close pairs and hence the II contribution.  The cross-correlation of well separated redshift bins, where the lensing is most efficient, enhances the GI contribution.  For this reason we cannot ignore the intrinsic ellipticity terms in equation~\ref{eqn:eeIIGIGG} when undertaking a tomographic weak lensing analysis and need to apply methods to mitigate the impact of this astrophysical contamination to the signal.  These mitigation strategies unfortunately limit the dramatic order-of-magnitude improvements initially anticipated from tomographic weak lensing analyses, but we still expect to find improved constraining power when implementing a tomographic analysis \citep{BK07}.

A series of different proposals to mitigate the impact of intrinsic alignments on weak lensing measurements have been made in the literature, and in some cases applied to data.   An optimal weighting scheme to down-weight physically close pairs was proposed by \citet{HH03} and applied to a weak lensing analysis of the COMBO-17 survey \citep[][see also King \& Schneider 2002]{HBH04}.   This method, however, only negates the II contamination and requires prior knowledge of the angular and redshift dependence of the intrinsic ellipticity correlation function in order to optimally analyse the data.  Motivated by our lack of knowledge in this area, \citet{JS08} proposed a nulling method where only the characteristic redshift dependence of the II and GI correlations in different combinations of redshift bins (see discussion above) are used to derive an optimal weighting scheme that, in an ideal case, nulls all II and GI contributions to the tomographic analysis.  This method, however, has been shown to significantly degrade cosmological parameter constraints \citep{JS09}.  

The main alternative to mitigating intrinsic alignment contamination by using different weighting schemes, is instead to use a model fitting approach, and it is this technique that we exploit in this paper.  This approach was first highlighted by \citet{King05} who demonstrated that a simultaneous model fitting analysis of finely binned tomographic data, using a sufficiently flexible parametrized model for the II and GI signals, allows for the marginalisation over the II and GI nuisance parameters  in the final cosmological analysis.   This idea was extended by \citet{BK07} to determine how the figure of merit for future cosmological surveys would degrade based on the flexibility of the II and GI model, or how much prior knowledge we are prepared to assume, and the accuracy of the photometric redshifts for the survey.   For a typical photometric redshift error of $\sigma_z \sim 0.05(1+z) $ the figure of merit for measuring dark energy would decrease by 20 to 50 per cent depending on the allowed flexibility of the model.  Similar conclusions were drawn by \citet{KTH08} and \citet{KT11} when investigating the loss of constraining power when intrinsic alignment modeling was marginalised over in a tomographic or fully three-dimensional weak lensing analysis, and more recently by \citet{Blazek12} who investigated the impact of marginalising over intrinsic alignments in a galaxy-galaxy lensing analysis.  The model fitting approach does, however, become more promising if the lensing data are analysed simultaneously with galaxy clustering data.  These extra data act to self-calibrate the II and GI signals \citep{Zhang10,JBSB10}.  \citet{Kirk10} present the first example of such a joint analysis, combining two-dimensional weak lensing data from the 100 square degree lensing survey \citep{JB07} and SDSS shear-shape clustering data \citep{RM06}.  A single parameter analytical model for intrinsic alignments from \citet{HS04} is used in this analysis, and we describe this `non-linear intrinsic alignment' model in more detail in Section~\ref{sec:NLA}. \citet{Huff} also use this model to remove GI contamination to their cosmic shear analysis of the SDSS-Stripe 82 survey, but they fix the contamination using a mean measurement of the total amplitude for the GI signal as determined by \citet{Hirata07}.  Finally,  \citet{Fu08} also mitigate the impact of GI on their cosmic shear analysis of the third year data from Canada-France-Hawaii Telescope Legacy Survey (CFHTLS), presenting a simultaneous two-dimensional weak lensing and intrinsic alignments model fitting method.  This analysis however uses an alternative model for the GI contamination, motivated by numerical simulations \citep{HeymansIA06}.  

A third way to account for the intrinsic alignment signals has been proposed, modifying the covariance matrix such that a marginalisation over possible functional forms of the II and GI power spectrum is permitted \citep{KT11}. This has the advantage in that all functional forms are explored, and does not require explicit estimation of any nuisance parameters, but makes the assumption that the variance of the intrinsic alignment functions is Gaussian.

The first combined weak lensing and photometric redshift analyses to directly detect the growth of structure, came from the COMBO-17 survey \citep{Bacon05,TKC17} and CFHTLS-Deep fields \citep{ES06}.   These analyses were followed by two independent tomographic analyses of the Hubble Space Telescope COSMOS survey \citep{RM_COSMOS, TS10}.  The areas of all these surveys were considered sufficiently small, all less than a few square degrees, such that any contributions from intrinsic alignments could be ignored in comparison to the large statistical errors.  \citet{TS10} did, however, attempt to mitigate any errors by removing all auto-correlated narrow redshift bins from their analysis to reduce the impact of the enhanced II signal in those bins.  In addition they removed all luminous red galaxies from their galaxy sample as the shapes of this type of galaxy have been shown to be the most strongly correlated with the local density field \citep{BJ11}.   As survey sizes grow and statistical errors decrease, it is not possible to ignore intrinsic alignments when analysing weak lensing in redshift bins.  The recent two-bin tomographic analysis of the 154 square degree CFHTLenS, presented in \citet{Benjamin2012}, therefore combines the strategies of \citet{Huff} and \citet{TS10} to mitigate intrinsic alignment contamination.  Using the linear tidal field intrinsic alignment model of \citet{HS04}, and following \citet{BK07} by fixing its amplitude to the observational constraints obtained by \citet{BTHD02}, they estimate the II and GI contamination to the cosmic shear measurement.  They then limit their analysis to two broad high redshift bins with photometric redshifts $0.5<z_{\rm ph}<0.85$ and $0.85<z_{\rm ph}<1.3$ such that any contamination from intrinsic alignments is expected to be no more than a few per cent.  This broad bin tomography measurement is also used to constrain parametrized modified gravity models in \citet{Simpson}.  In this paper we use the same CFHTLenS data, presenting the first tomographic weak lensing analysis to apply a full model fitting approach to mitigate the impact of intrinsic alignment contamination on shear correlation functions.   A fully three-dimensional weak lensing analysis \citep{KHM11} is applied to the same CFHTLenS data in \citet{Kitching2012}. 

This paper is set out as follows.  In Section~\ref{sec:data} we describe CFHTLenS and the auxiliary data sets used in this analysis.  We outline our methodology and chosen intrinsic alignment model in Section~\ref{sec:method}, additionally describing how our tomographic analysis is constrained by our requirements on the accuracy of the covariance matrix estimated from N-body lensing simulations.  We present our results in Section~\ref{sec:res}, comparing joint parameter constraints from different combinations of CFHTLenS data with the cosmic microwave background data from WMAP7, baryon acoustic oscillations data from BOSS, and a prior on the Hubble constant from the HST distance ladder.  In Section~\ref{sec:IAredblue} we focus on the constraints that can be placed on the amplitude of the intrinsic alignment  signal for early-type and late-type galaxies with this type of cosmological parameter analysis, with concluding remarks in Section~\ref{sec:conc}.

\section{The Canada-France-Hawaii Telescope Lensing Survey}
\label{sec:data}
The Canada-France-Hawaii Telescope Lensing Survey (CFHTLenS) is a 154 square degree deep multi-colour $u^*g'r'i'z'$ survey optimised for weak lensing analyses, observed as part of the CFHT Legacy Survey (CFHTLS) on the 3.6m Canada-France-Hawaii telescope.  The data span four distinct contiguous fields:  W1 ($\sim 63.8$ square degrees), W2 ($\sim 22.6$ square degrees), W3 ($\sim 44.2$ square degrees) and W4 ($\sim 23.3$ square degrees). The CFHTLenS analysis of these data presents the current state-of-the-art in weak lensing data processing with {\sc THELI} \citep{Erben12}, shear measurement with lens{\em fit} \citep{Miller2012}, photometric redshift measurement from PSF-matched photometry  \citep{HH12} using the Bayesian photometric redshift code {\sc BPZ} \citep{BPZ}, and a stringent systematic error analysis \citep{syspaper}.   The resulting galaxy catalogue that we use in this analysis includes a shear measurement $\epsilon_{\rm obs}$ with an inverse variance weight ${\rm w}$ and a photometric redshift estimate $z_{\rm BPZ}$ with a probability distribution $P(z)$ and best-fit photometric galaxy type $T_{\rm BPZ}$.   We apply the galaxy size and signal-to-noise dependent shear calibration corrections described in \citet{Miller2012} and \citet{syspaper}, and only use the subset of 75 per cent of the survey data that have been verified as science-ready and free of significant systematic errors.  This has been demonstrated through a series of rigorous cosmology-insensitive tests on both the shear and photometric redshifts measurements, in combination \citep[see][for the full details]{syspaper}.  \citet{Benjamin2012} also use a cross-correlation analysis to verify the accuracy of the measured redshift distributions $P(z)$ when the galaxy sample is limited to those galaxies with a most probable photometric redshift estimate between $0.2 < z_{\rm BPZ} < 1.3$.  In light of these analyses, that demonstrate the robustness of these data to systematic errors, we do not present any further systematic error analyses in this work, referring the reader to \citet{syspaper}, and references therein.  For the redshift selection $0.2 < z_{\rm BPZ} < 1.3$, the galaxy sample has a weighted mean redshift of $\bar{z} = 0.75$, and a weighted median redshift of $z_{\rm m} = 0.70$, as determined from the weighted sum of the $P(z)$.   The effective weighted galaxy number density, in this redshift range, is $n_{\rm eff} =11$ galaxies per square arcmin.   

\subsection{Auxiliary cosmological data}
In this analysis we present joint cosmological parameter constraints by combining our tomographic weak lensing analysis of CFHTLenS with up to three complementary data sets to break parameter degeneracies.    We include the temperature and temperature-polarisation cosmic microwave background power spectra from the {\it Wilkinson Microwave Anisotropy Probe} \citep[][hereafter referred to as WMAP7]{WMAP7data}.    We incorporate the measurement of baryonic acoustic oscillations using the {\it Baryon Oscillation Spectroscopic Survey} data from the ninth data release of the {\it Sloan Digital Sky Survey} \citep[][hereafter referred to as BOSS]{BOSS}.  We adopt their primary reconstructed distance constraint $D_V(z=0.57) / r_s = 13.67 \pm 0.22$.   Here $r_s$ is the sound horizon at the baryon drag epoch, and $D_V(z=0.57)$ is the volume element at a redshift $z=0.57$ which depends on angular diameter distances and the Hubble parameter $H(z)$.   This constraint is found to be in excellent agreement with measurements of the distance-redshift relation from Type Ia supernovae \citep{SNLS3,Suzuki}.  As baryon acoustic oscillations and supernova are probing similar geometric properties of the Universe, using current supernova data in combination with WMAP7 and BOSS yields little to no improvement for the majority of cosmological parameters that we constrain in this analysis \citep[see][for more details]{BOSS}.  We therefore do not include Type Ia supernovae constraints.  We do however include a Gaussian prior on the Hubble constant, $H_0=73.8 \pm 2.4 \, {\rm km \, s}^{-1} {\rm Mpc}^{-1}$,  which combines constraints from local supernovae, Cepheid variables, and the megamaser at the centre of NGC 4258 \citep[][hereafter referred to as R11]{R11}.

\section{Method}
\label{sec:method}
In this section we review the theory and measurement of weak lensing in tomographic redshift bins, discuss the non-linear intrinsic alignment model that we adopt for this analysis, and present our method to estimate the covariance matrix error from the \citet{Clone2012} suite of high resolution N-body lensing simulations.    We focus on a real-space shear correlation function analysis in this paper, presenting a fully 3D spherical harmonic analysis in \citet{Kitching2012}.  We conclude this section describing the properties of the chosen tomographic redshift bins and the Population Monte Carlo method that we use to determine cosmological parameter constraints from the data.

\subsection{Weak Lensing Tomography}
\label{sec:theory}
Weak gravitational lensing by large-scale structure induces weak correlations between the observed shapes of distant galaxies.   We parametrize galaxy shape through the complex galaxy ellipticity $\epsilon = \epsilon_1 + {\rm i} \epsilon_2$.  The simple relationship between ellipticity and shear, given in equation~\ref{eqn:egamma}, holds for weak shear, when the ellipticity for a perfect ellipse with an axial ratio $\beta$ and orientation $\phi$, is defined as
\be
\left(
\begin{array}{c}
\epsilon_1 \\
\epsilon_2
\end{array}
\left)
= \frac{\beta-1}{\beta+1}
\right(
\begin{array}{c}
\cos 2\phi \\
\sin 2 \phi
\end{array}
\right) \, .
\label{eqn:e1e2}
\ee
There are a range of different two-point statistics that have been proposed to extract weak lensing information from the data \citep[see][for a comprehensive discussion of the relationship between these statistics]{SchvWKM02,COSEBIS}.  These statistics, however, all stem from a base measurement of the observed angular two-point correlation function $\hat{\xi}_{\pm}$ which can be estimated from two redshift bins, $i$ and $j$, from the data as follows:
\be
\hat{\xi}_{\pm}^{ij}(\theta) = \frac{\sum {\rm w}_a {\rm w}_b \left[ \epsilon_{t}^i (\bm{x_a}) \epsilon_{t}^j (\bm{x_b}) \, \pm \, \epsilon_\times^i (\bm{x_a}) \epsilon_\times^j (\bm{x_b})
\right]}{
\sum {\rm w}_a {\rm w}_b } \, .
\label{eqn:xipm_est}
\ee
Here the weighted sum,  using inverse variance weights ${\rm w}$, is taken over galaxy pairs with angular separation $|\bm{x_a - x_b}|=\theta$.   The tangential and cross ellipticity parameters $\epsilon_{{\rm t},\times}$ are the ellipticity parameters in equation~(\ref{eqn:e1e2}) rotated into the reference frame joining each pair of correlated objects.     In this paper we only focus on this statistic, referring the reader to \citet{Kilbinger2012} who present a non-tomographic analysis of CFHTLenS using a wide range of different two-point statistics.  This analysis demonstrates that for CFHTLenS, the cosmological parameter constraints are insensitive to the two-point statistic adopted for the analysis. 

The two-point shear correlation function $\xi_{\pm}(\theta)_{\rm GG}$ is related to the underlying non-linear matter power spectrum $P_\delta$ that we wish to probe, with
\be
\xi_{\pm}^{ij}(\theta)_{\rm GG} = \frac{1}{2\pi}\int d\ell \,\ell \,P_\kappa^{ij}(\ell) \, J_{\pm}(\ell \theta) \, , 
\label{eqn:xiGG}
\ee
where $J_\pm (\ell \theta)$ is the zeroth (for $\xi_+$) and fourth (for $\xi_- $) order Bessel function of the first kind. $P_\kappa(\ell)$ is the convergence power spectrum at angular wave number $\ell$ 
\be 
P_\kappa^{ij}(\ell) = \int_0^{w_{\rm H}} dw \, 
\frac{q_i(w)q_j(w)}{[f_K(w)]^2} \, P_\delta \left( \frac{\ell}{f_K(w)},w \right),
\label{eqn:Pkappa} 
\ee
where $f_K(w)$ is the angular diameter distance out to the comoving radial distance, $w$, and 
$w_H$ is the horizon distance.  The lensing efficiency function, $q_{i}(w)$, for a redshift bin $i$, is given by
\be
q_i(w) = \frac{3 H_0^2 \Omega_{\rm m}}{2c^2} \frac{f_K(w)}{a(w)}\int_w^{w_{\rm H}}\, dw'\ n_i(w') 
\frac{f_K(w'-w)}{f_K(w')}, 
\label{eqn:qk} 
\ee
where $n_i(w)dw$ is the effective number of galaxies in $dw$ in redshift bin $i$, normalized so that $\int n_i(w)dw = 1$.  $a(w)$ is the dimensionless scale factor, 
$H_0$ is the Hubble parameter and $\Omega_{\rm m}$ the matter density parameter at $z=0$.  For more details see \citet{Bible} and references therein.

\subsection{Non-Linear Intrinsic Alignment Model}
\label{sec:NLA}
In this paper we adopt the non-linear intrinsic alignment model developed by \citet{BK07} to parametrize the contribution of intrinsic alignments to our tomographic shear measurement.  This model is a simplified version of the linear tidal field alignment model derived analytically by \citet{HS04}, based on the earlier work of \citet{CKB01}.  \citet{BK07} choose to make one key addition to this model by replacing the linear matter power spectrum with a non-linear power spectrum, hence the name, non-linear intrinsic alignment model.  This modification to the original model was motivated by comparisons of the model predictions to measurements of intrinsic alignments from data \citep{RM06}, and simulations \citep{HBH04}, and the desire to make the linear tidal field alignment model more realistic on small scales.  This model has since been adopted by several observational analyses \citep{Kirk10,BJ11,RM11}, as it has the useful property that with only a single parameter $A$, both the II and GI contribution to the shear correlation function can be predicted.   The non-linear intrinsic alignment II and GI power spectra are related to the non-linear matter power spectrum as, 
\ba
P_{\rm II}(k,z) =  F^2(z) P_\delta(k,z) & P_{\rm GI}(k,z) =  F(z) P_\delta(k,z) \, ,
\label{eqn:PIIGI} 
\ea
where the redshift and cosmology-dependent modification to the power spectrum is given by
\be 
F(z) = - A C_1 \rho_{\rm crit} \frac{\Omega_{\rm m}}{D(z)} \, .
\label{eqn:Fz} 
\ee
Here $\rho_{\rm crit}$ is the critical density at $z=0$ and $D(z)$ is the linear growth factor normalized to unity today.  We follow \citet{BJ11} by parameterizing the amplitude of $F(z)$ with a free dimensionless amplitude parameter, $A$,  and a fixed normalization constant $C_1 = 5 \times 10^{-14} \, h^{-2} M_\odot^{-1} {\rm Mpc}^3$.  The value of $C_1$ is chosen so that the model matches the observational results of \citet{BTHD02} such that the fiducial model for our analysis will assume $A=1$.  In this case the GI term is negative and acts to decrease the overall signal.  The II term, however, is always positive, independent of the sign of $A$, and acts to increase the overall signal.  We note that $F(z)$ differs from \citet{BK07}, as we incorporate the redshift dependent corrections to the linear tidal field alignment model reported in \citet{HS04erratum,BJ11}.

The II and GI contributions to the observed two-point correlation function are analogous to the GG contribution from equation~\ref{eqn:xiGG},
\be
\xi_{\pm}^{ij}(\theta)_{\rm II, GI} = \frac{1}{2\pi}\int d\ell \,\ell \,C_{\rm II,GI}^{ij}(\ell) \, J_{\pm}(\ell \theta) \, , 
\label{eqn:xiIIGI}
\ee
with the convergence power spectrum $P_\kappa$ replaced by the projected GI power spectrum $C_{\rm GI}$ or projected II power spectrum $C_{\rm II}$, 
\be 
C_{\rm II}^{ij}(\ell) = \int_0^{w_{\rm H}} dw \, 
\frac{n_i(w)n_j(w)}{[f_K(w)]^2} \, P_{\rm II} \left( \frac{\ell}{f_K(w)},w \right),
\label{eqn:CII} 
\ee
\be 
C_{\rm GI}^{ij}(\ell) = \int_0^{w_{\rm H}} dw \, 
\frac{q_i(w)n_j(w) + n_i(w)q_j(w) }{[f_K(w)]^2} \, P_{\rm GI} \left( \frac{\ell}{f_K(w)},w \right),
\label{eqn:CGI} 
\ee
where the projection takes into account the effective number of galaxies $n(w)$, and, in the case of GI correlations,  the lensing efficiency $q_{i}(w)$ (equation~\ref{eqn:qk}).   Consider two non-overlapping distinct redshift bins, such that $n_i(w)n_j(w) = 0$ for all $w$.   As the II term comes from physically close galaxy pairs, we find for these non-overlapping bins $C_{\rm II}=0$,  as expected.  The GI term comes from the correlation of background shear with foreground intrinsic ellipticities.  For the same two non-overlapping bins, with the mean redshift of bin $j$ greater than the mean redshift of bin $i$, we see that only the $n_i(w)q_j(w)$ term in the projection is non-zero, again as expected for GI.  In practice, however, we will find that statistical and catastrophic errors in photometric redshift estimation will result in some level of overlap between all the bins, so we expect some contribution from II and GI between all our tomographic bin combinations. 

Following from equation~\ref{eqn:eeIIGIGG}, we can now relate the observed two-point ellipticity correlation function (equation~\ref{eqn:xipm_est}) to the two-point shear correlation function that we wish to measure (GG term, see equation~\ref{eqn:xiGG}) and the two types of intrinsic alignment contamination (II and GI) that we wish to marginalise over in our weak lensing analysis,
\be
\hat{\xi}_{\pm}^{ij}(\theta) = \xi_{\pm}^{ij}(\theta)_{\rm II} +\xi_{\pm}^{ij}(\theta)_{\rm GI} +\xi_{\pm}^{ij}(\theta)_{\rm GG} \, .
\ee

Before we continue to discuss our analysis technique to extract cosmological parameters from the two-point ellipticity correlation function, we should pause to assess how realistic the adopted non-linear intrinsic alignment model is.   As noted in \citet{Hirata07},  the \citet{CKB01} model of linear galaxy alignment has no sound grounding in theory,  in contrast to the analytical models of angular momentum correlations, and hence galaxy alignment, from tidal torque theory \citep[see the review by][and references therein]{Schafer09}.  It does however provide a good match to simulations and observations, \citep[see for example][]{Hirata07,BK07} and is therefore currently the favoured model.  Equation~\ref{eqn:Fz} assumes that the amplitude $A$ of the intrinsic galaxy alignment signal is independent of redshift, and that even at the smallest physical scales, the II and GI power spectra follow the same scale-dependent evolution as the underlying matter power spectrum.    Both these assumptions are clearly over-simplifications and a truly conservative tomographic analysis should therefore implement a more flexible intrinsic alignment model, allowing for scale and redshift dependent perturbations around the non-linear intrinsic alignment model as proposed by \citet{BK07}.  We choose not to take this approach, however, motivated by the null measurement of intrinsic alignments using a sample of galaxies that are most similar to the CFHTLenS galaxy population, at the median redshift of  CFHTLenS, by \citet{RM11}.  This observational analysis
places an upper bound of $0.03$ ($2\sigma$ limit) on the size of the systematic error induced on a measurement of $\sigma_8$ when neglecting intrinsic alignments in a CFHTLenS-like survey. We can therefore conclude that the systematic error introduced by the presence of intrinsic alignments in CFHTLenS is likely to be sufficiently small that any additional systematic error we introduce by using an inflexible single-parameter intrinsic alignment model will be insignificant compared to our statistical errors (which we show in Section~\ref{sec:wlcomp} result in a $1\sigma$ error of $\pm 0.04$ on $\sigma_8$ at fixed $\Omega_m$, when intrinsic alignments are marginalised over).  This rationale will no longer hold for future surveys, however, where the large survey areas will reduce the statistical errors, highlighting the pressing need for improved models and observations of intrinsic galaxy alignments. 

\subsection{Covariance Matrix Estimation}
\label{sec:covest}
One of the major challenges with a tomographic lensing analysis of the type that we use in this analysis, is the construction of a sufficiently accurate covariance matrix that accounts for the significant levels of correlation between both the angular and redshift binned data.  For ${\rm N}_{\rm t}$ tomographic redshift bins, ${\rm N}_\theta$ angular scales and considering both the $\xi_+^{ij}$ and $\xi_-^{ij}$ components of the shear correlation function between redshift bins $i$ and $j$, we have a total number of data points $p$ given by
\be
p = {\rm N}_\theta \, {\rm N}_{\rm t} ({\rm N}_{\rm t} +1) \, .
\label{eqn:p}
\ee
It is both the data vector, $\bm{D}(p)$, and the inverse of the corresponding $p \times p$ covariance matrix, $\mathbfss{C}^{-1}$, that we use in the cosmological parameter likelihood analysis.  In this section we show that it is the covariance matrix estimation that limits the maximum number of data points that we can accurately analyse, motivating and justifying our choice of $p=210$ in Section~\ref{sec:nofz}.

There are a number of methods that can be used to estimate a covariance matrix.  Bootstrap and Jackknife techniques can be used to estimate the data covariance directly from the survey \citep{Wallstats}, and various de-noising techniques can be applied to reduce the impact of noise bias when the matrix is inverted \citep[see for example][]{Norberg09}.  Analytical functions can be derived for random Gaussian convergence fields \citep{SchvWKM02,KS04} and scale-dependent non-Gaussian corrections can be applied to those analytical functions as determined through comparison to N-body lensing simulations \citep{Semboloni07,Satofit}.  Analytical functions can also be derived for log-normal convergence fields or using a halo model, which are found to be a significantly better approximation compared to a Gaussian field.  These analytical functions, however, still require calibration using N-body simulations \citep{Hilbert11,KTJ12}.  With sufficient numbers of 3D N-body lensing simulations, the covariance can also be estimated directly from the simulations.  These can be populated to emulate the survey geometry, masks, redshift distribution, galaxy number density and intrinsic ellipticity and shape measurement noise for the survey \citep{Millraytrace,Vafaei10,Alinasims,Clone2012}.  

There are advantages and disadvantages to each method.  Direct estimation of the covariance matrix from the survey provides a cosmology-independent, but noisy estimate.    In the inversion, the covariance matrix can therefore become unstable.  Furthermore, this method is only suitable when the survey contains many independent lines of sight, adequate to evaluate sampling variance errors which dominate at large scales.  As such, this direct estimation is only suitable for very large area surveys.  Analytical functions are precise, but approximate.  They can be calculated for any cosmology, but require cosmology and scale-dependent corrections, for angular scales $\theta \ls 20$ arcmin, calibrated with N-body simulations, to account for the Gaussian or log-normal approximations made in their calculation.   N-body simulations are costly, and for the real-space correlation analysis presented in this paper, the requirements on simulation resolution and size is demanding.  Low particle resolution in the simulation results in an artificial lack of power that propagates to all scales with real-space statistics.   The finite simulation box size also truncates density perturbations on scales larger than the box length which results in a suppression of power in the large-scale real-space dark matter halo correlation function measured from the simulations \citep{PK06}.    High resolution, large area, 3D lensing simulations that are suitable for our real-space analysis are therefore rare, and for a fixed cosmology \citep{Millraytrace,Clone2012}.  

In this analysis we estimate a covariance matrix from the three-dimensional N-body numerical lensing simulations of \citet{Clone2012}.  Light cones are formed from line of sight integration through independent dark matter particle simulations.  The simulated cosmology matches the 5-year WMAP flat $\Lambda$CDM cosmology constraints from \citet{WMAP5}.  The $1024^3$ particle simulations have a box size of $147.0 \, {\rm h}^{-1} {\rm Mpc}$ or $231.1 \, {\rm h}^{-1} {\rm Mpc}$, depending on the redshift of the simulation.  The boxes are grafted in projection such that each high resolution line of sight simulation has a real space resolution of 0.2 arcmin in the shear field, spanning 12.84 square degrees sampled at 26 redshift slices between $0<z<3$.    The two-point shear statistics measured in real-space from the simulations closely matches the theoretical predictions of the input cosmology from $0.5 \ls \theta \ls 40$ arcmin scales at all redshifts \citep{Clone2012}.   We therefore limit our tomographic analysis to these angular scales where we can obtain an accurate estimate of the covariance matrix.

We populate each simulation by mapping 12.84 square degree sections of the galaxy spatial distribution from the survey on to the simulated shear field, and the corresponding redshift distribution and galaxy weights.  With this method the survey masks and effective source density are correctly replicated in the simulations.   We create a redshift realization of each line of sight, by sampling a galaxy redshift $z$ at random from each galaxy $P(z)$ in the survey.  We verified that our results were insensitive to this realization method by repeating the measurement of the covariance for different random seeds.  A galaxy shear $\gamma$ is then assigned at that redshift by linearly interpolating between the shear output at discrete redshifts by the lensing simulations.   Intrinsic ellipticity and shape measurement noise is added by randomly sampling from a zero-mean Gaussian of width $\sigma_e = 0.279$ per ellipticity component as measured from the data\footnote{\citet{Miller2012} show that the true intrinsic galaxy ellipticity distribution is not well approximated by a Gaussian distribution.  For the purposes of evaluating the dominant shot noise component of the covariance matrix, however, we argue that a Gaussian approximation is sufficient.  The width of the Gaussian model used is calculated directly from the data such that the variance of the data and simulated ellipticity distributions are the same, even if the shapes of the distributions differ. For the sub-dominant mix-term part of the covariance, however, which arises from correlations between galaxy ellipticities and cosmic shear, our Gaussian approximation could lead to a mild underestimate of the covariance and this should be investigated in future work.}.  Note that we found no significant variation in the shot noise $\sigma_e$ when measured as a function of photometric redshift within our high confidence redshift range $0.2 < z_{\rm BPZ} < 1.3$.  This was determined by binning the data into photometric redshift slices of width $\Delta z_{\rm BPZ} = 0.1$, where $\sigma_e$ for each bin was found to lie within two standard deviations of the mean of the full sample.

\citet{Clone2012} create a total of 184 fully independent N-body simulations.  In order to gain a sufficient number of realizations such that the correction for correlated noise in the inverse covariance matrix will not bias our results (see the discussion in Section~\ref{sec:invcov}), we are therefore required to split each of the simulations into 9 semi-independent simulations.  Each simulation then spans 1.4 square degrees such that the centres are separated by more than twice the largest angular scale probed by this analysis.  This method to increase the number of simulations by sub-division, also employed by \citet{TS10}, yields $n_\mu = 1656$ semi-independent line of sight simulations.   In Section~\ref{sec:wlcomp} we show that the low level of correlation expected between each group of 9 simulations, introduced by splitting each fully independent simulation into sub-fields, does not impact significantly on our results.  
The data vector $\bm{D}^\mu$ of the $p$ components of $\xi^{ij}_{\pm}(\theta)$, is then measured from each simulation line of sight $n_\mu$ and the covariance matrix is estimated,
\be
\hat{\mathbfss{C}}(a,b) = \frac{1}{A_s n_\mu} \sum_{\mu=1}^{n_\mu}  ( D^{\mu}_a - {\bar D}_a ) ( D^{\mu}_b - {\bar D}_b ) \, ,
\label{eqn:cov}
\ee
where $\bar{\bm{D}}$ is the average of the data measured over the total $n_\mu$ simulations.   The area scaling term $A_s$ accounts for the difference in area between one line of sight in the simulation and the CFHTLenS survey area used in the analysis \citep{SchvWKM02}.  

The covariance matrix we derive with this method ignores the impact of intrinsic galaxy alignments.  As our fiducial $A=1$ intrinsic alignment model is expected to lower the amplitude of the observed two-point correlation function only at the few per cent level, a GG-only covariance matrix is therefore expected to only slightly over-estimate our errors\footnote{In the analysis that follows we find $A$ to be consistent with zero for the full galaxy sample and a sample of late-type galaxies, supporting this strategy.  For early-type galaxies we find a significant value for $A$.  In this case, however, shot noise dominates the covariance matrix as the early-type galaxy sample comprises only 20 per cent of the full galaxy population.  In this case it is therefore again a reasonable strategy to ignore the impact of intrinsic galaxy alignments on the covariance matrix.}.  We therefore conclude a GG-only covariance is sufficient for our analysis supported by the findings of \citet{GrocuttPhD} who analysed the covariance matrix derived from simulated II+GI+GG Gaussian fields  created using the method of \citet{BrownBattye}.  For future tomographic analyses we will populate the N-body lensing simulations used in this analysis with an intrinsic ellipticity component \citep[see for example][]{HeymansIA06}.

The covariance matrix we derive with this method is dependent on the fixed simulation cosmology.  \citet{Eifler09} investigate the impact of including a varying cosmology covariance matrix in a cosmic shear analysis, concluding that fixing the cosmology of the covariance matrix has a non-negligible impact on the size of the likelihood contours.  This is an effect that becomes less pronounced as the statistical power of a survey increases.  \citet{Kilbinger2012} therefore present a detailed analysis of the impact of including a varying cosmology covariance matrix in the analysis of the 2D cosmic shear measurement for the CFHTLenS data.  They also present a novel method to combine the advantages of all the covariance matrix estimation methods to create a cosmology dependent covariance matrix.  On scales $\theta<30$ arcmins, the non-Gaussian sampling variance is estimated from the N-body lensing simulations of \citet{Clone2012}, as described above, but setting $\sigma_e = 0$.  A varying cosmology estimate is then obtained by scaling the data vector $\bm{D}^{\mu}$ by the ratio of the desired cosmology to the fixed simulation cosmology.   For the larger angular scales, where the finite box size causes power to be underestimated in the simulations, a Gaussian analytic model is used, which has been shown to be a good approximation on these scales  \citep{Semboloni07,Satofit}.  The shot-noise term of the covariance matrix is also estimated analytically \citep{SchvWKM02} and the cosmology variation in the mixed term, which accounts for the covariance between the cosmic shear and shot noise, is modeled using the \citet{Eifler09} fitting formulae.  The conclusion of this detailed covariance matrix analysis of \citet{Kilbinger2012} is that, for a CFHTLenS-like survey, the impact of using a fixed cosmology covariance matrix, or a varying cosmology covariance matrix is marginal, particularly when CFHTLenS results are used in combination with other cosmology surveys.    \citet{Kilbinger2012} also show a good agreement between the diagonal components of the covariance matrix with a Jackknife estimate of the covariance from the data.  We therefore conclude that using the \citet{Clone2012} simulations to produce a fixed cosmology covariance matrix, as estimated using Equation~\ref{eqn:cov}, is sufficiently accurate for our purposes.

\subsubsection{Inverse Covariance Estimation}
\label{sec:invcov}
For the likelihood analysis of the CFHTLenS data, we require the inverse of the covariance matrix.   Whilst we consider our measurement of $\hat{\mathbfss{C}}$ from N-body lensing simulations to be an unbiased estimator of the true covariance matrix $\mathbfss{C}$, it will have an associated measurement noise from averaging over a finite number of semi-independent realizations $n_\mu$.  This measurement noise means that $\hat{\mathbfss{C}}^{-1}$ is not an unbiased estimate of the true inverse covariance matrix \citep{HartlapCor}.  Assuming Gaussian measurement errors on $\hat{\mathbfss{C}}$,  an unbiased estimate of the true inverse covariance matrix is derived in \citet{Anderson}, where $\mathbfss{C}^{-1} =  \alpha_{\rm A} \, \hat{\mathbfss{C}}^{-1}$ and
\be
\alpha_{\rm A} =  \frac{ n_\mu -p -2}{n_\mu-1} \, .
\label{eqn:anderson}
\ee
\citet{HartlapCor} shows that for $p/n_\mu < 0.8$, this correction produces an unbiased estimate of the inverse covariance matrix $\mathbfss{C}^{-1}$ and we use this correction in our analysis.

Unfortunately, even though the use of the \citet{Anderson} correction creates an inverse covariance that is now unbiased in the mean, there are still associated measurement errors on the individual components of $\hat{\mathbfss{C}}^{-1}$ which now become boosted by a factor $\alpha_{\rm A}$. \citet{HartlapCor} show that boosting this measurement noise in the inverse covariance, impacts on the size of the confidence regions resulting from a likelihood analysis.  They find that the area of the confidence regions can erroneously grow by up to 30 per cent as $p/n_\mu \rightarrow 0.8$.  
The consequence is that for high values of $p/n_\mu$, parameter constraints appear less significant than they truly are.  If we require that the inclusion of the \citet{Anderson} correction  does not increase the area of our Bayesian confidence regions by more than $\sim 5$ per cent, \citet{HartlapCor} estimate a limit of $p/n_\mu \ls 0.12$ \citep[see also][for similar conclusions]{GrocuttPhD}. We note that the tomographic analysis of the COSMOS survey by \citet{TS10} estimated the covariance matrix from the Millennium Simulation with $p/n_\mu = 0.55$.   Whilst well within the requirements for an unbiased \citet{Anderson} correction,  \citet{HartlapCor} predicts such a correction would increase the area of the confidence regions in a likelihood analysis by $\sim 12$ per cent.  As such the \citet{TS10} cosmological parameter constraint errors are expected to be slightly overestimated.

As we have a fixed number of N-body lensing simulations which determines $n_\mu$, it is the accuracy that we require for the inverse covariance matrix $\mathbfss{C}^{-1}$ that sets the maximum number of data points $p$ in our analysis.  The number of tomographic bins ${\rm N}_{\rm t}$ and angular scales ${\rm N}_\theta$ is therefore set by the number of N-body simulations that we have at our disposal.  For $p/n_\mu \ls 0.12$, and $n_\mu=1656$, (see Section~\ref{sec:covest}), we should therefore limit our analysis to $p \ls 200$.

\subsection{Tomographic analysis and redshift distributions}
\label{sec:nofz}

In a tomographic weak lensing analysis there is always a choice to be made for the number of tomographic redshift bins,  ${\rm N}_{\rm t}$, and the number of scales probed, in our case angular scales, ${\rm N}_\theta$.    As the number of redshift and angular bins is increased, the amount of information increases.  A saturation limit is eventually reached beyond which the data points become so correlated that the extra information gained with each incremental increase in the number of bins becomes marginal.   With an unlimited number of N-body lensing simulations from which to make an unbiased covariance matrix estimate,  the optimal number of tomographic bins will depend on the photometric redshift accuracy of the survey, and the method by which the contamination from intrinsic galaxy alignments is mitigated in the analysis.   \citet{BK07} show that for a survey with a photometric redshift scatter of $\sigma_z = 0.05 (1+z)$, using ${\rm N}_{\rm t} \sim 8$ brings the cosmological parameter constraints to within 20 per cent of the best attainable with a fully 3D approach.     This is in contrast to the conclusions of earlier cosmic-shear only optimizations, which found ${\rm N}_{\rm t} \sim 3$ to be optimal \citep{SKS04, MaHuHut}.  This difference indicates the importance of using finely binned tomographic redshift slices when mitigating intrinsic alignment effects. \citet{GrocuttPhD} investigate the dependence of cosmological parameter constraints when varying the number of tomographic redshift bins,  ${\rm N}_{\rm t}$, and the number of angular scales probed, ${\rm N}_\theta$, simultaneously.  A non-linear intrinsic alignment model was assumed for the II and GI contamination (see Section~\ref{sec:NLA}).  In this analysis the cosmological parameter constraints were found to be less sensitive to increases in ${\rm N}_\theta$, in comparison to increases in ${\rm N}_{\rm t}$.  This is expected for the single-parameter non-linear intrinsic alignment model,  as the cosmic shear, GG, and non-linear intrinsic alignment II and GI power spectrum, vary smoothly with scale and the relative amplitude between the II, GI and GG power for each redshift bin is fixed as a function of scale.    As the number of data points $p$ scales as ${\rm N}_{\rm t}({\rm N}_{\rm t}+1)$, however, even small increases in ${\rm N}_{\rm t}$ can quickly lead to an unstable covariance matrix.

\begin{table}
\caption{Tomographic redshift bin selection.  Galaxies are selected based on their maximum posterior photometric redshift estimate $z_{\rm BPZ}$.  The median redshift $z_{\rm m}$ and mean redshift $\bar{z}$ for each bin is calculated from the effective redshift distribution as measured by the weighted sum of the photometric redshift error distributions $P(z)$.   }
\centering
\begin{tabular}{c c c c}
\hline
Bin & $z_{\rm BPZ}$  & $z_{\rm m}$ & $\bar{z}$ \\ \hline\hline
  1& $ 0.20\, - \, 0.39$ & 0.28& 0.36\\ \hline
  2& $ 0.39\, - \, 0.58$ & 0.48& 0.50\\ \hline
  3& $ 0.58\, - \, 0.72$ & 0.62& 0.68\\ \hline
  4& $ 0.72\, - \, 0.86$ & 0.82& 0.87\\ \hline
  5& $ 0.86\, - \, 1.02$ & 0.93& 1.00\\ \hline
  6& $ 1.02\, - \, 1.30$ & 1.12& 1.16\\ \hline
\end{tabular}
\label{tab:zbins}
\end{table}

Motivated by the findings of  \citet{BK07} and \citet{GrocuttPhD}, and with the limitation that the total number of data points $p \ls 200$ (see Section~\ref{sec:invcov}), we choose to use $N_{\rm t} = 6$ redshift bins and $N_\theta = 5$ angular bins such that our total number of data points $p=210$. 
The angular range is chosen to be spaced equally in $\log(\theta)$ between $1.5<\theta<35$ arcmin, where the maximum angular scale is determined by the limitations of the N-body lensing simulations used to determine the covariance matrix. 
We select the $N_{\rm t}=6$ redshift bins to span our high confidence redshift range $0.2 < z_{\rm BPZ} < 1.3$ such that the effective angular number density of galaxies in each redshift bin is equal.  The effective number density includes the shear measurement weights ${\rm w}$ such that the intrinsic ellipticity noise in each bin is equal.  This choice is in contrast to a cosmic shear signal-to-noise optimised redshift bin selection which would lead to much broader bins at low redshift.  Such optimization is undesirable for our purposes, as it is the lowest redshift bins where the presence of intrinsic alignments is most prominent.  Table~\ref{tab:zbins} lists the resulting redshift selection for each tomographic bin.  The median redshift $z_{\rm m}$ and mean redshift $\bar{z}$ is calculated from the effective redshift distribution as measured by the weighted sum of the photometric error distributions $P(z)$.  These error distributions extend out to $z_{\rm BPZ} = 3.5$ which skews the mean redshift measurement, relative to the median, particularly in the lowest redshift bin. 

\begin{figure}
   \centering
   \includegraphics[width=2.7in, angle=270]{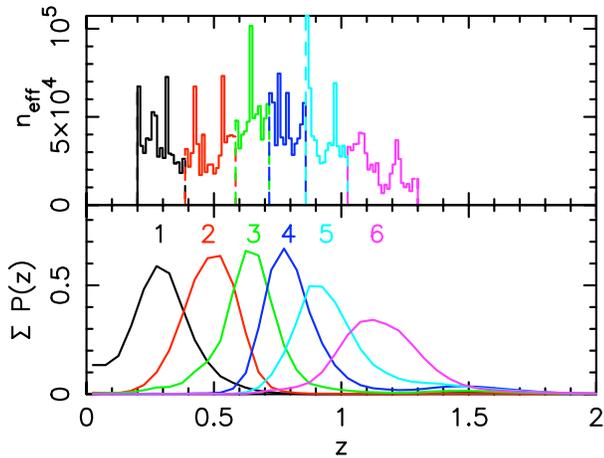} 
   \caption{Tomographic redshift distribution.  The upper panel shows the effective weighted number of galaxies as a function of their maximum posterior photometric redshift estimate, separated into six tomographic bins between $0.2<z_{\rm BPZ}<1.3$.  The effective weighted number of galaxies in each redshift bin is constant.  The lower panel shows the redshift distribution for each selected bin as estimated from the weighted sum of the photometric redshift probability distributions $P(z)$.}
   \label{fig:nofz}
\end{figure}

Figure~\ref{fig:nofz} compares the effective redshift distribution for each tomographic bin as determined from the maximum posterior redshift $z_{\rm BPZ}$ (upper panel) and by the weighted sum of the photometric redshift error distributions $P(z)$ (lower panel).   In order to highlight the differences between the redshift distributions measured from these two different estimators, the binning in the upper panel is chosen to be significantly finer than the typical CFHTLenS photometric redshift error $\sigma_z \sim 0.04(1+z)$ \citep{HH12}.  The fine structure revealed by this binning therefore illustrates redshift focusing effects arising from the photometric redshift measurement, not true physical structures.  Accurate measurements of $P(z)$ for each galaxy allows us to fully account for these focussing effects, in addition to overlapping redshift distributions and catastrophic redshift outliers in our analysis \citep[see][for detailed analysis of the $P(z)$ used in this analysis]{Benjamin2012}.  For our intrinsic alignment analysis it is particularly important to quantify the degree of overlap between redshift bins as the II term is only significant for physically close galaxy pairs.  It is therefore the summed $P(z)$ redshift distributions displayed in the lower panel of Figure~\ref{fig:nofz} that we use in this analysis.

\subsection{Population Monte Carlo Sampling likelihood analysis method}
\label{sec:PMC}

In this study we perform a Bayesian likelihood analysis of CFHTLenS and the auxiliary data, discussed in Section~\ref{sec:data}, to constrain the parameters of a range of cosmological models.  To calculate the likelihood values we use the Population Monte Carlo sampling software {\sc CosmoPMC}\footnote{CosmoPMC: www.cosmopmc.info} \citep{CosmoPMC}, modified to include an optional simultaneous fit of cosmic shear and the intrinsic alignment model outlined in Section~\ref{sec:NLA}.  Future releases of this software package will include this option.  The Population Monte Carlo method is described in \citet{PMC} along with a comparison to the more standard Markov-Chain Monte Carlo method for cosmological parameter estimation.   We also refer the reader to a detailed discussion of the {\sc CosmoPMC} analysis of 2D CFHTLenS cosmic shear data in \citet{Kilbinger2012} for further information about the methodology.

We assume a matter power spectrum derived from the \citet{EH08} transfer function with a \citet{Rob} non-linear correction.  For dark energy cosmologies, where the equation of state of dark energy parameter, $w_0 \neq -1$, a modulation of the non-linear power is required \citep{MTC06} which we apply using of the scaling correction from \citet{TS10,icosmo}.  The \citet{Rob} halo-model prescription for the non-linear correction has been calibrated on numerical simulations and shown to be accurate to between 5 per cent and 10 per cent over a wide range of $k$ scales \citep{Eifler11} and found to be of sufficient accuracy for the statistical power of CFHTLenS \citep{V12}.  Whilst our assumed transfer function includes baryonic oscillations on large scales,  we are unable to include the uncertain effects of baryons on small physical scales.   \citet{ESbaryons} present an analysis of cosmological hydrodynamic simulations to quantify the effect of baryon physics on the weak gravitational lensing shear signal, using a range of different baryonic feedback models.  For the $\xi_+$ angular scales we use  we would expect baryons to induce at most a $\sim 10$ per cent decrease in the signal relative to a dark matter only Universe, in the mid-to-high redshift tomographic bins where our highest signal-to-noise measurements are made.  This is assuming the `AGN feedback' model which leads to the largest changes in the matter power spectrum of the simulations
that were studied by \citet{ESbaryons}, where we note that this scenario is the one that matches observed gas fractions in groups.    In the cosmological analysis that follows, we present an additional conservative analysis where the tomographic data most susceptible to significant errors caused by baryonic or non-linear effects are removed \citep[see][for further discussion]{Benjamin2012}.    If significant errors exist, however, the inclusion and marginalisation over the intrinsic alignment amplitude $A$ in our analysis, which modulates the amplitude of the observed shear power spectrum, should work to some extent, to reduce the impact of these effects in addition to mitigating contamination by intrinsic galaxy alignments.

We use {\sc CosmoPMC} to analyse CFHTLenS and WMAP7 independently.  For the combined results with BOSS and our assumed $H_0$ prior from R11, we importance-sample the WMAP7-only likelihood chain, multiplying each sample point with the CFHTLenS, BOSS and R11 posterior probability.   For our CFHTLenS-only flat $\Lambda$CDM analysis we limit our parameter set to the matter density parameter, $\Omega_{\rm m}$, the amplitude of the matter power spectrum controlled by $\sigma_8$, the baryon density parameter $\Omega_b$, the Hubble parameter $h$, and the power spectrum spectral index $n_{\rm{s}}$.  With WMAP7 we also include into the parameter set the reionisation optical depth $\tau$, the Sunyaev-Zel'dovich template amplitude $A_{\rm SZ}$, and the primordial amplitude of the matter perturbations $\Delta_{\rm{R}}^2$, from which we derive $\sigma_8$.  The  equation of state of dark energy parameter, $w_0$ and dark energy density parameter $\Omega_{\rm de}$ are also included for non-flat or non-$\Lambda$ cosmological models.   We use flat priors throughout which are broad enough to cover the full $3\sigma$ posterior distribution in each parameter direction for each combination of data.  Throughout the paper we quote and plot 68 per cent and 95 per cent Bayesian confidence or credibility regions.  These regions contain 68 per cent and 95 per cent of the posterior probability determined from the multi-dimensional distribution of points from the PMC parameter sample.  All figures showing the resulting joint-constraints on two parameters, are marginalised over the multi-dimensional parameter space that is not shown.

\section{Results}
\label{sec:res}

\begin{figure*}
   \centering
   \includegraphics[width=5.5in, angle=270]{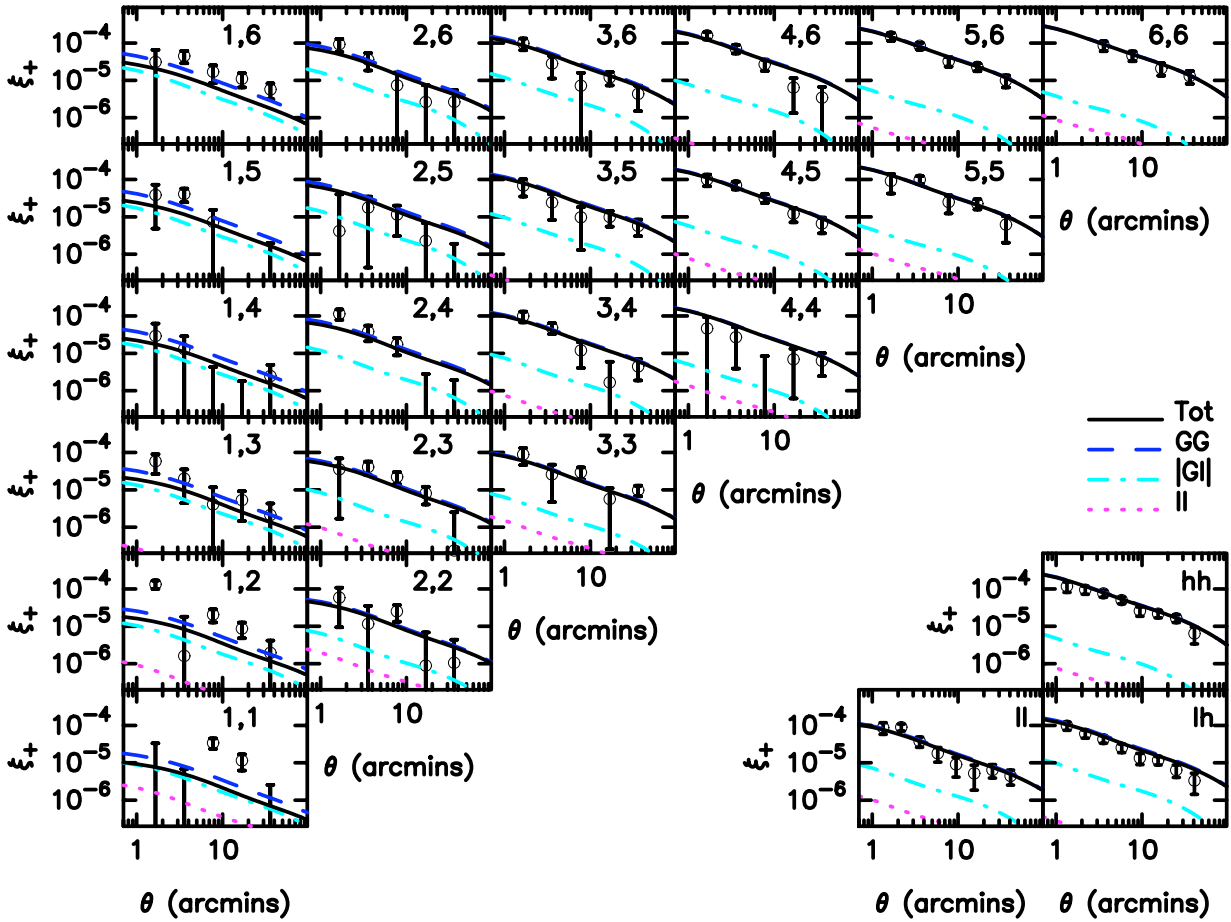} 
   \caption{The observed two-point correlation function $\hat{\xi}^{ij}_+(\theta)$.  The panels show the different $ij$ redshift bin combinations, ordered with increasing redshift bin $i$ from left to right, and increasing redshift bin $j$ from lower to upper.  Refer to table~\ref{tab:zbins} for the redshift ranges of each tomographic bin.  The errors are estimated from an analysis of N-body lensing simulations as discussed in Section~\ref{sec:covest}.   The theoretical curves show our fiducial total GG+GI+II signal as a solid line.  When distinguishable from the total, the GG only signal is shown dashed.  The magnitude of the GI signal is shown dot-dashed (our fiducial GI model has a negative anti-correlated signal) and the II signal is shown dotted, where the amplitude is more than $10^{-7}$.  The results of the broad two-bin tomographic analysis of \citet{Benjamin2012} are shown in the lower right corner.}
   \label{fig:xip}
\end{figure*}

Figure~\ref{fig:xip} presents the observed two-point correlation function $\hat{\xi}^{ij}_+(\theta)$ for every tomographic bin combination in our chosen six redshift bin analysis.  With $N_{\rm t}$ tomographic bins, there are $N_{\rm t}(N_{\rm t}+1)/2$ independent combinations, or 21 combinations in our case.  The panels show the different $ij$ bin combinations, ordered with increasing redshift bin $i$ from left to right, and increasing redshift bin $j$ from lower to upper, where the redshift distributions of each bin are shown and tabulated in Section~\ref{sec:nofz}.  The auto-correlated bins lie along the diagonal.    The data points are calculated using the shear correlation function estimator in Equation~\ref{eqn:xipm_est}, correlating pairs of galaxies within the full mosaic catalogue for each of the four CFHTLS fields.  The measurements from each field are then combined using a weighted average, where the field weight is given by the effective number of galaxy pairs in each angular bin.  Note that the results for each $ij$ bin from each field were found to be noisy but consistent \citep[see][for measurements of the higher signal-to-noise 2D shear correlation function for each CFHTLS field]{Kilbinger2012}.  The errors, which include sample variance, are estimated from an analysis of N-body lensing simulations as discussed in Section~\ref{sec:covest}.   We remind the reader that the data are highly correlated, particularly in the low redshift bins.  The theoretical curves show our fiducial WMAP7 best-fit cosmological parameter model, with an $A=1$ non-linear intrinsic alignment model, to be a good fit to the data.  A possible exception to this are data from tomographic bin combinations that include the lowest redshift bin, which we discuss further in Section~\ref{sec:vis}.  The individual components are shown; GG (dashed), GI (dot-dashed) and II (dotted) models with the total GG+GI+II shown as a solid line.  For comparison we also show the results of the broad two-bin tomographic analysis of \citet{Benjamin2012} in the lower right corner to demonstrate the low-level of II and GI contamination expected for this high redshift selected analysis.

\subsection{Tomographic Data Visualization}
\label{sec:vis}

With 21 tomographic bin combinations, two statistics $\hat{\xi}^{ij}_+(\theta)$ and $\hat{\xi}^{ij}_-(\theta)$, and 5 angular scales, we have a total of $p=210$ data points, half of which are shown in Figure~\ref{fig:xip}.  In the cosmological parameter constraints that follow, it is this large data vector, and a correspondingly large covariance matrix, that we use in the likelihood analysis.  
Purely for improving the visualization of this large data set, however, we propose the following method to compress the data, motivated by the different methods of \citet{RM_COSMOS} and \citet{TS10}.   

To compress angular scales, we first calculate a WMAP7 cosmology GG-only theory model $\xi^{ij}_{\rm fid}$ for each redshift bin combination $ij$ and each statistic ($+/-$).  We then define a free parameter $\alpha_\pm^{ij}$ which allows the overall amplitude of the model to vary, but keeps the angular dependence fixed.  The best-fitting amplitude $\alpha_\pm^{ij}$ is then found from a $\chi^2$ minimization of $\alpha_\pm^{ij}\xi^{ij}_{\rm fid}(\theta)$  to the shear correlation functions measured at 5 angular scales in each $ij$ bin and each statistic.   A best-fitting value of $\alpha_\pm^{ij} = 1$ implies the data in bin $ij$ are well-fit by a WMAP7 GG-only cosmology.  Following \citet{TS10}, each bin is then assigned a single value of $\alpha^{ij}\, \hat{\xi}^{ij}_{\rm fid}(\theta=1')$ which can be interpreted as the amplitude of the two-point shear correlation function measured in bin $ij$ at an angular scale of $\theta=1$ arcmin.

To compress the information in the redshift bin combination, we calculate the lensing efficiency function $q_i(w)$ (equation~\ref{eqn:qk}) for each redshift bin $i$, and then determine the peak redshift $z_{\rm peak}$ of the combined lensing sensitivity $q_i(w) q_j(w)$ for each redshift bin $ij$ combination.  This peak redshift locates the epoch that is the most efficient at lensing the two galaxy samples in the redshift bin combination $ij$, but we note that these distributions are very broad, particularly for the redshift bins with a significant fraction of catastrophic outliers in the photometric redshift distribution (see Figure~\ref{fig:nofz}).  

\begin{figure}
   \centering
   \includegraphics[width=3.0in, angle=270]{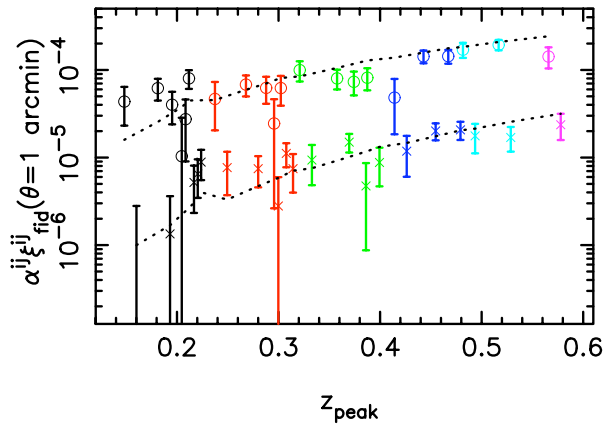} 
   \caption{Compressed CFHTLenS tomographic data where each point represents a different tomographic bin combination $ij$ as indicated by $z_{\rm peak}$, the peak redshift of the lensing efficiency for that bin combination.  The best-fitting amplitude $\alpha^{ij}$ of the data relative to a fixed fiducial GG-only cosmology model is shown, multiplied by the fiducial model at $\theta = 1$ arcmin for $\xi_+$ (circles) and $\xi_-$ (crosses, offset along the $z_{\rm peak}$ axis for clarity).  The error bars show the $1\sigma$ constraints on the fit.  The data can be compared to the fiducial GG-only model, shown dotted.  Note that the colour of the points follow the same colour-scheme as Figure~\ref{fig:nofz}, and indicates the lower redshift bin that is used for each point.}   
   \label{fig:datacomp}
\end{figure}

Figure~\ref{fig:datacomp} shows the resulting compressed 21 data points for each statistic, $\xi_+$ (circles) and $\xi_-$ (crosses), plotting $\alpha^{ij}\, \hat{\xi}^{ij}_{\rm fid}(\theta=1')$ against $z_{\rm peak}$.  This can be compared to the fiducial cosmology prediction (shown dotted,  by setting $\alpha=1$).  Note that the relatively high fraction of catastrophic redshift outliers in the lowest redshift bin impacts on the expected signal measured from redshift bin combinations that include this bin.  The expected increase in signal, as $z_{\rm peak}$ increases, is therefore not smooth.  This can be seen in the theoretical curve in Figure~\ref{fig:datacomp} which displays a slight kink at $z_{\rm peak} = 0.22$.   To recover $\alpha^{ij}$ from this figure, one simply divides the value of each data point by the value of the fiducial model, shown dotted, at that $z_{\rm peak}$.  Consistent values for $\alpha^{ij}$ are measured from both the $\xi_+$ and $\xi_-$ statistic.
We find a signal that rises as the peak redshift of the lensing efficiency function increases; the more large scale structure the light from our background galaxies propagates through, the stronger the lensing effect.  In general, the data are well-fit by the WMAP7 GG-only fiducial model, but we do see an indication of an excess signal at low redshifts where, for a fixed angular scale, the smaller physical scales probed are more likely to be contaminated by the intrinsic galaxy alignment signal.  This is however also the regime where the analysis is most affected by catastrophic outliers in our photometric redshift distribution.  Based on the cross-correlation analysis of \citet{Benjamin2012} we expect these errors to be accounted for by our use of photometric redshift distributions $P(z)$.  In \citet{syspaper}, we also show that the catalogues used in this analysis present no significant redshift-dependent systematic bias when tested with a cosmology-insensitive galaxy-galaxy lensing analysis.  This gives us confidence in the robustness of our results at all redshifts.  We note that in order to make this visualization of the data, the different redshift bin combinations and the $\xi_+$ and $\xi_-$ statistics are considered to be uncorrelated.  The plotted $1\sigma$ errors on $\alpha$ are therefore underestimated but we re-iterate at this point that this data compression is purely for visualization purposes and it is not used in any of the cosmological parameter constraints that follow.

\subsection{Comparison of parameter constraints from weak lensing in a flat $\Lambda$CDM cosmology}
\label{sec:wlcomp}
\begin{figure}
   \centering
   \includegraphics[width=3.4in, angle=0]{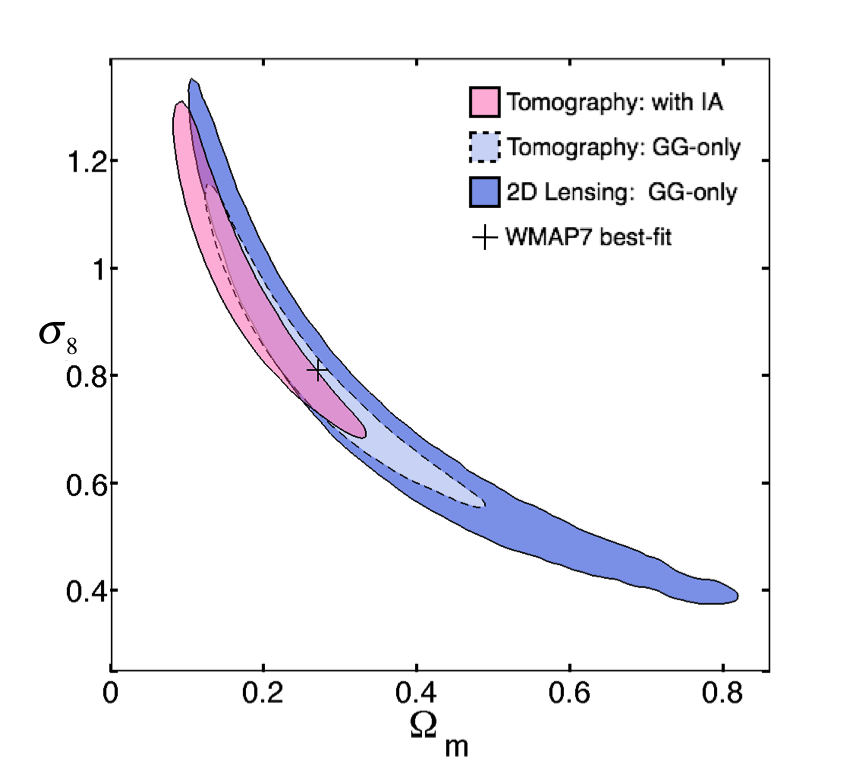} 
   \caption{Flat $\Lambda$CDM parameter constraints (68 per cent confidence) on the amplitude of the matter power spectrum controlled by $\sigma_8$ and the matter density parameter $\Omega_{\rm m}$ from CFHTLenS-only, comparing three cases:  2D weak lensing (blue) and 6-bin tomographic lensing where intrinsic alignments are assumed to be zero (pale blue) and are marginalised over (pink). For reference, the black cross shows the corresponding best-fit values from WMAP7 \citep{WMAP7}.   }
   \label{fig:s8om_GG_2D_comp}
\end{figure}

The measurement of cosmological weak lensing alone is most sensitive to the overall amplitude of the matter power spectrum.  This depends on a degenerate combination of the clustering amplitude $\sigma_8$ and the matter density parameter $\Omega_{\rm m}$, and it is therefore in this parameter space that we choose to compare the constraints we find from weak lensing alone using different analysis techniques.  We limit this comparison to flat $\Lambda$CDM cosmologies.  Figure~\ref{fig:s8om_GG_2D_comp} compares three cases.  In blue we show the 68 per cent Bayesian confidence limits from a 2D weak lensing analysis of CFHTLenS, limited to the same angular scales as our tomography analysis with $\theta<35$ arcmin.  This can be compared to the  68 per cent constraints from our 6-bin $\xi_{\pm}$ tomographic lensing measurement when intrinsic alignments are assumed to be zero (pale blue) and when the amplitude of the intrinsic alignment model is allowed to be a free parameter and is marginalised over (pink).  All three measurements are consistent and can be compared to the best-fit WMAP7 results shown as a black cross for reference.  

\begin{table}
\renewcommand{\arraystretch}{1.4} 
\caption{Constraints orthogonal to the $\sigma_8-\Omega_{\rm m}$ degeneracy direction for a number of different types of lensing analyses: 2D weak lensing to $\theta<350$ arcmin \citep{Kilbinger2012} and to $\theta<35$ arcmin,  2-bin tomographic lensing \citep{Benjamin2012}, and 6-bin tomographic lensing where intrinsic alignments are assumed to be zero or are marginalised over (our primary result indicated in bold).  We also present constraints for two conservative analyses to test the covariance matrix and our sensitivity to potential error in the assumed non-linear correction to the matter power spectrum (see text for more detail).}
\centering
\begin{tabular}{lcc} \hline
Data & $\alpha$  & $ \sigma_8(\Omega_{\rm m}/0.27)^\alpha$ \\ \hline\hline
2D Lensing:   & & \\
$\theta<350$' \citep{Kilbinger2012}& $0.59 \pm 0.02$ & $0.787 \pm 0.032$ \\
$\theta<35$' & $0.61 \pm 0.02$ & $0.791^{+0.052}_{-0.066}$ \\ \hline\hline
2-bin tomography:& & \\ 
$\theta<40$' \citep{Benjamin2012}&  $0.55 \pm 0.02$  & $ 0.771 \pm 0.040$ \\\hline\hline
6-bin tomography (all $\theta<35$'):  & & \\ 
$A=0$ & $0.52 \pm 0.02$ & $0.783^{+0.024}_{- 0.032}$\\
$\bm{A}$ \bf{marginalised} & $\bm{0.46 \pm 0.02}$ & $ \bm{0.774^{+0.032}_{- 0.041}}$\\ 
$n_\mu = 736$ covariance matrix & $0.48 \pm 0.03 $ & $0.768^{+0.032}_{- 0.041} $ \\
Low $\theta$ scales removed & $0.45 \pm 0.03$ & $0.774^{+0.038}_{-0.057}$ \\ \hline
\end{tabular}
\label{tab:s8omres} 
\end{table}

Table~\ref{tab:s8omres} lists the parameter constraints, for the three cases shown in Figure~\ref{fig:s8om_GG_2D_comp}, on the combination 
$\sigma_8(\Omega_{\rm m}/0.27)^\alpha$.  The parameter $\alpha$ is derived from a fit to the likelihood surface to determine the direction that is orthogonal to the $\sigma_8-\Omega_{\rm m}$ degeneracy direction.  These results can be compared to the 2D CFHTLenS constraints from \citet{Kilbinger2012}, where large angular scales were included in the analysis, and a 2-bin tomography analysis from \citet{Benjamin2012}, limited to the same angular scales considered in this analysis.  We find excellent agreement between the cosmological results from the different analyses, indicating that ignoring intrinsic alignment contamination in \citet{Kilbinger2012} and \citet{Benjamin2012} did not introduce any significant bias in their results. Differences in the values of the $\alpha$ parameter arise from a number of factors. The strength of the lensing signal is modulated by $\Omega_{\rm m}$ in a manner which is sensitive to both the source redshift distribution and the angular scales under consideration \citep{BVWM97}. Furthermore the degeneracy contours are not perfectly represented by a power law, so the value of $\alpha$ is not our key interest here.

Focusing first on the constraints from tomography and 2D lensing limited to the same angular scales but ignoring intrinsic alignments (shown blue and pale blue in Figure~\ref{fig:s8om_GG_2D_comp}), we find close to a factor of two improvement in the constraint on $\sigma_8(\Omega_{\rm m}/0.27)^\alpha$, in addition to an improvement in degeneracy breaking between $\sigma_8$ and $\Omega_{\rm m}$, when tomographic bins are considered.   Unfortunately,  however, our tomographic analysis is limited by the extent of the N-body simulations used to determine our covariance matrix, which forces us to lose the large angular scales considered in the 2D analysis from \citet{Kilbinger2012}.  Comparing the constraints from tomography limited to $\theta < 35$ arcmin, with 2D lensing out to $\theta = 350$ arcmins, we find similar constraints on $\sigma_8(\Omega_{\rm m}/0.27)^\alpha$.  This demonstrates how, in this parameter space, the large angular scales are adding as much information in a 2D Lensing analysis as the additional redshift bins add in a tomographic analysis of the same data.  This motivates future work to remove the current angular limitations imposed by the tomographic covariance matrix estimation method that we use in this analysis.  

Adding in the intrinsic alignment model, and hence an additional free parameter $A$, broadens the parameter constraints, as expected, reducing the constraining power on $\sigma_8(\Omega_{\rm m}/0.27)^\alpha$ by roughly 30 per cent compared to a GG-only analysis.  For $\Omega_{\rm m}=0.27$, however, we find very little difference in the best-fit value of $\sigma_8$ which changes by 0.01.  Larger deviations between the two analyses are however seen for higher values of $\Omega_{\rm m}$.  Figure~\ref{fig:s8om_GG_2D_comp} shows that the 68 per cent confidence region shifts to slightly lower $\Omega_{\rm m}$ and higher $\sigma_8$.  For a fixed $\Omega_{\rm m}$, the resulting best-fit $\sigma_8$ is therefore slightly lower when intrinsic alignments are marginalised over.  This behaviour is unexpected for a conventional intrinsic alignment model where the negative GI signal dominates over the positive II signal such that the total GG+GI+II signal observed is less than the GG signal alone.  For the fiducial $A=1$ intrinsic alignment model, a GG-only analysis would therefore underestimate $\sigma_8$ for a fixed $\Omega_{\rm m}$.  For this CFHTLenS-only analysis, however, we instead find a preference for a negative value of $A=-1.60^{+ 1.33}_{-1.94}$, and hence the GG-only analysis prefers higher values of $\sigma_8$.  We explore and discuss this result in more detail in Section~\ref{sec:IAredblue} but re-iterate that the differences we have commented upon here are well within our $2\sigma$ errors and are therefore not significant.

Finally we perform two conservative analyses to ensure the robustness of our results, the constraints from which are reported in the lower two rows of table~\ref{tab:s8omres} .  The first is to compare constraints when we use a covariance matrix constructed from $n_\mu = 736$ semi-independent lines of sight (where each fully independent N-body lensing simulation is split into 4 sub-simulations)  instead of the standard $n_\mu= 1656$ analysis that we use throughout this paper.  The excellent agreement between the results from the two estimates of the covariance matrix verifies that the low level of correlation expected between each group of 9 or 4 simulations, introduced by splitting each fully independent simulation into sub-fields, does not impact significantly on our results.  It also demonstrates that the \citet{Anderson} inverse covariance de-biasing correction (equation~\ref{eqn:anderson}), is sufficiently accurate for our analysis.  The second conservative analysis is to remove the angular scales where uncertainty in the accuracy of the non-linear correction to the power spectrum could bias our results.  We select these angular scales by calculating a WMAP7 cosmology theoretical prediction for $\xi^{ij}_\pm(\theta)$ assuming two different non-linear corrections, where we boost and decrease the \cite{Rob} non-linear correction to the power spectrum by $\pm 7$ per cent.  Note that we chose the value of 7 per cent from the average error over the range of $k$ scales tested in \citet{Eifler11}.   For angular scales where more than a 10 per cent difference is found in the expected signal, between these two different non-linear correction regimes, we remove these scales from our analysis.  As the $\xi_-$ statistic probes significantly smaller $k$ scales compared to the $\xi_+$ statistic, at a fixed $\theta$, we cut more $\xi_-$ data than $\xi_+$ \citep[see][for further discussion]{Benjamin2012}.   For $\xi_+$, our requirement for less than a 10 per cent deviation corresponds to the removal of data with $\theta \ls 3$ arcmin for tomographic bin combinations including bins 1 and 2.  For $\xi_-$, this corresponds to removing data with $\theta \ls 30$ arcmin for tomographic bin combinations including bins 1, 2, 3 and 4, and data with $\theta \ls 16$ arcmin for tomographic bin combinations including bins 5 and 6.  Applying these cuts in angular scale results in a final data vector of length $p=120$.  As the $\xi_\pm$ statistic is an integral over many $k$ scales weighted by $J_0$ and $J_4$ Bessel functions, one cannot directly relate the limits we place on $\theta$, to limits on $k$.  We note, however, that as these cuts do preferentially remove the smallest physical $k$ scales where the theoretical prediction to the power spectrum is expected to be most impacted by baryonic feedback effects.  This conservative analysis to test the non-linear correction therefore also works as a mitigation strategy to avoid uncertain baryon feedback errors.  For this conservative analysis we find no change in the best-fit measurement of $\sigma_8(\Omega_{\rm m}/0.27)^\alpha$, but a reduction in the constraining power by roughly 20 per cent (see the `Low $\theta$ scales removed' row in table~\ref{tab:s8omres}).   
We also lose roughly 20 per cent of the constraining power on the intrinsic alignment amplitude $A$ with this conservative analysis.  As the best-fit value for $\sigma_8(\Omega_{\rm m}/0.27)^\alpha$ remains unchanged, we can conclude that the inclusion of small-scale data does not introduce any significant bias in our results.  Furthermore, as our focus for this analysis is the mitigation of intrinsic galaxy alignments,  which are most tightly constrained by the low-redshift bins preferentially cut with this type of conservative analysis, the CFHTLenS results that follow include the full angular scale range shown in Figure~\ref{fig:xip}.

\subsection{Joint Cosmological Parameter constraints}
\label{sec:cosmoparam}
\begin{figure}
   \centering
   \includegraphics[width=3.4in, angle=0]{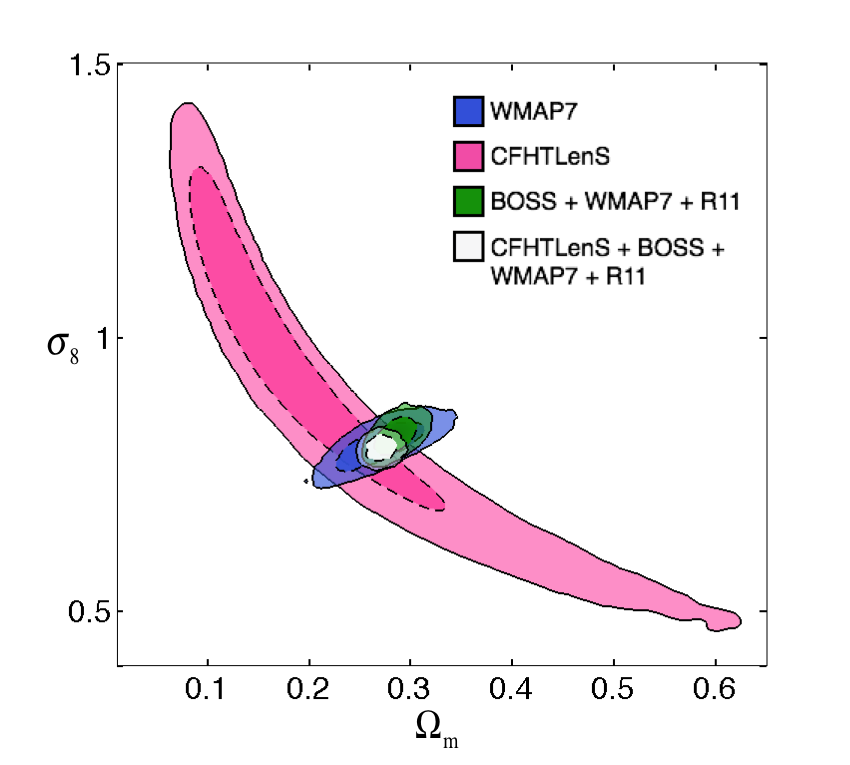} 
   \caption{Flat $\Lambda$CDM joint parameter constraints (68 and 95 per cent confidence) on the amplitude of the matter power spectrum controlled by $\sigma_8$ and the matter density parameter $\Omega_{\rm m}$ from CFHTLenS-only (pink), WMAP7-only (blue), BOSS combined with WMAP7 and R11 (green), and CFHTLenS combined with BOSS, WMAP7 and R11 (white).}   
   \label{fig:s8om}
\end{figure}

We present joint cosmological parameter constraints from CFHTLenS combined with WMAP7, BOSS and R11 for four cosmological models testing flat and curved $\Lambda$CDM and $w$CDM cosmologies.  Table~\ref{tab:cospar} lists the best-fit 68 per cent confidence limits for our cosmological parameter set for the combination of CFHTLenS and WMAP7 (first line for each parameter), CFHTLenS, WMAP7 and R11 (second line for each parameter) and for CFHTLenS, WMAP7, BOSS and R11 (third line for each parameter).  For comparison the figures in this section also show constraints for WMAP7-only and WMAP7 combined with BOSS and R11.  We refer the reader to \citet{WMAP7} and \citet{BOSS} for tabulated cosmological parameter constraints for the non-CFHTLenS combination of data sets shown, noting that we find good agreement with their tabulated constraints.  We also refer the reader to \citet{Kilbinger2012} for CFHTLenS-only parameter constraints for the curved and $w$CDM cosmological models tested in this section.  Whilst CFHTLenS currently represents the most cosmologically constraining weak lensing survey, it spans only 154 square degrees and is therefore not expected to have significant constraining power when considered alone.   This is demonstrated in Figure~\ref{fig:s8om} which compares parameter constraints in the $\sigma_8 -\Omega_{\rm m}$ plane for a flat $\Lambda$CDM cosmology.  The wide constraints from CFHTLenS-only are shown in pink (note the inner 68 per cent confidence region was shown in pink in Figure~\ref{fig:s8om_GG_2D_comp}), in comparison to the tight constraints from WMAP7-only (blue).  The power of lensing, however arises from its ability to break degeneracies in this parameter space owing to the orthogonal degeneracy directions.  BOSS combined with WMAP7 and R11 is shown green and when CFHTLenS is added in combination with BOSS, WMAP7 and R11 (white) we find the combined confidence region decreases in area by nearly a factor of two.  As we will show in this section, the tomographic lensing information presented in this analysis is therefore very powerful when used in combination with auxiliary data sets.

\begin{table*}
\renewcommand{\arraystretch}{1.2} 
\caption{Joint cosmological parameter constraints for four models, testing flat and curved $\Lambda$CDM and $w$CDM cosmologies. The first line for each parameter lists the constraints from CFHTLenS combined with WMAP7 and R11.  The second line lists the constraints from CFHTLenS combined with WMAP7, BOSS and R11. Deduced parameters are indicated with $^\star$. 
}
\centering
\begin{tabular}{llllll}\\ \hline
Parameter   & flat $\Lambda$CDM & flat $w$CDM  &  curved $\Lambda$CDM  & curved $w$CDM  & Data \\ \hline\hline
$\Omega_m$&
$  0.255^{+  0.014}_{-  0.014}$ &
$  0.256^{+  0.111}_{-  0.073}$ &
$  0.255^{+  0.028}_{-  0.023}$ &
$  0.214^{+  0.161}_{-  0.049}$ &CFHTLenS + WMAP7
\\ 
&
$  0.250^{+  0.012}_{-  0.012}$ &
$  0.242^{+  0.020}_{-  0.014}$ &
$  0.248^{+  0.014}_{-  0.013}$ &
$  0.243^{+  0.020}_{-  0.014}$ &CFHTLenS + WMAP7 + R11
\\ 
&
$  0.271^{+  0.010}_{-  0.009}$ &
$  0.269^{+  0.018}_{-  0.015}$ &
$  0.275^{+  0.011}_{-  0.010}$ &
$  0.247^{+  0.021}_{-  0.018}$ &CFHTLenS + WMAP7 + R11 + BOSS
\\ \hline
$\sigma_8^*$&
$  0.794^{+  0.016}_{-  0.017}$ &
$  0.81^{+  0.10}_{-  0.10}$ & 
$  0.805^{+  0.028}_{-  0.029}$ &
$  0.871^{+  0.076}_{-  0.125}$ &CFHTLenS + WMAP7
\\ 
&
$  0.795^{+  0.016}_{-  0.018}$ &
$  0.810^{+  0.030}_{-  0.027}$ &
$  0.813^{+  0.021}_{-  0.024}$ &
$  0.819^{+  0.028}_{-  0.032}$ &CFHTLenS + WMAP7 + R11
\\ 
&
$  0.799^{+  0.014}_{-  0.016}$ &
$  0.800^{+  0.030}_{-  0.025}$ &
$  0.791^{+  0.017}_{-  0.019}$ &
$  0.826^{+  0.026}_{-  0.031}$ &CFHTLenS + WMAP7 + R11 + BOSS
\\ \hline
$A$&
$   -1.18^{+    0.96}_{-    1.17}$ &
$   -1.4^{+    1.2}_{-    1.9}$ & 
$   -0.84^{+    0.97}_{-    1.21}$ &
$   -1.7^{+    1.4}_{-    2.0}$ &CFHTLenS + WMAP7 
\\ 
&
$   -1.37^{+    0.96}_{-    1.21}$ &
$   -1.3^{+    1.0}_{-    1.2}$ & 
$   -0.91^{+    0.94}_{-    1.04}$ &
$   -0.85^{+    0.89}_{-    1.16}$ &CFHTLenS + WMAP7 + R11
\\ 
&
$   -0.48^{+    0.75}_{-    0.87}$ &
$   -0.51^{+    0.82}_{-    0.84}$ &
$   -0.31^{+    0.70}_{-    0.86}$ &
$   -0.32^{+    0.70}_{-    1.04}$ &CFHTLenS + WMAP7 + R11 + BOSS
\\ \hline
$w_0$&
$-1$ &                                                                                                                                                                                                  
$   -1.05^{+    0.33}_{-    0.34}$ &
$-1$ &                                                                                                                                                                                                  
$   -1.18^{+    0.36}_{-    0.22}$ &CFHTLenS + WMAP7
\\ 
&
$-1$ &                                                                                                                                                                                                  
$   -1.06^{+    0.08}_{-    0.07}$ &
$-1$ &                                                                                                                                                                                                  
$   -1.04^{+    0.11}_{-    0.12}$ &CFHTLenS + WMAP7 + R11
\\ 
&
$-1$ &                                                                                                                                                                                                  
$   -1.02^{+    0.09}_{-    0.09}$ &
$-1$ &                                                                                                                                                                                                  
$   -1.19^{+    0.14}_{-    0.11}$ &CFHTLenS + WMAP7 + R11 + BOSS
\\ \hline
$\Omega_{\rm de}$&
$1-\Omega_m$ &                                                                                                                                                                                          
$1-\Omega_m$ &                                                                                                                                                                                          
$  0.743^{+  0.029}_{-  0.025}$ &
$  0.782^{+  0.161}_{-  0.050}$ &CFHTLenS + WMAP7
\\ 
&
$1-\Omega_m$ &                                                                                                                                                                                          
$1-\Omega_m$ &                                                                                                                                                                                          
$  0.747^{+  0.015}_{-  0.014}$ &
$  0.753^{+  0.022}_{-  0.016}$ &CFHTLenS + WMAP7 + R11
\\ 
&
$1-\Omega_m$ &                                                                                                                                                                                          
$1-\Omega_m$ &                                                                                                                                                                                          
$  0.730^{+  0.012}_{-  0.011}$ &
$  0.762^{+  0.021}_{-  0.019}$ &CFHTLenS + WMAP7 + R11 + BOSS
\\ \hline
$\Omega_{\rm K}^*$&
$0$ &                                                                                                                                                                                                   
$0$ &                                                                                                                                                                                                   
$  0.002^{+  0.008}_{-  0.009}$ &
$  0.004^{+  0.006}_{-  0.008}$ &CFHTLenS + WMAP7
\\ 
&
$0$ &                                                                                                                                                                                                   
$0$ &                                                                                                                                                                                                   
$  0.005^{+  0.005}_{-  0.005}$ &
$  0.004^{+  0.008}_{-  0.007}$ &CFHTLenS + WMAP7 + R11
\\ 
&
$0$ &                                                                                                                                                                                                   
$0$ &                                                                                                                                                                                                   
$ -0.004^{+  0.004}_{-  0.004}$ &
$ -0.009^{+  0.005}_{-  0.004}$ &CFHTLenS + WMAP7 + R11 + BOSS
\\ \hline
$h$&
$  0.717^{+  0.016}_{-  0.015}$ &
$  0.74^{+  0.14}_{-  0.12}$ & 
$  0.724^{+  0.042}_{-  0.041}$ &
$  0.82^{+  0.11}_{-  0.16}$ &CFHTLenS + WMAP7 
\\ 
&
$  0.723^{+  0.013}_{-  0.013}$ &
$  0.738^{+  0.023}_{-  0.026}$ &
$  0.734^{+  0.022}_{-  0.020}$ &
$  0.741^{+  0.022}_{-  0.024}$ &CFHTLenS + WMAP7 + R11
\\ 
&
$  0.702^{+  0.010}_{-  0.010}$ &
$  0.706^{+  0.023}_{-  0.020}$ &
$  0.691^{+  0.014}_{-  0.011}$ &
$  0.724^{+  0.023}_{-  0.027}$ &CFHTLenS + WMAP7 + R11 + BOSS
\\ \hline
$\Omega_b$&
$  0.0437^{+  0.0014}_{-  0.0014}$ &
$  0.044^{+  0.020}_{-  0.012}$ & 
$  0.0431^{+  0.0057}_{-  0.0041}$ &
$  0.0358^{+  0.0282}_{-  0.0086}$ &CFHTLenS + WMAP7
\\ 
&
$  0.0433^{+  0.0012}_{-  0.0013}$ &
$  0.0414^{+  0.0032}_{-  0.0026}$ &
$  0.0417^{+  0.0027}_{-  0.0023}$ &
$  0.0409^{+  0.0032}_{-  0.0024}$ &CFHTLenS + WMAP7 + R11
\\ 
&
$  0.0453^{+  0.0009}_{-  0.0011}$ &
$  0.0450^{+  0.0037}_{-  0.0029}$ &
$  0.0470^{+  0.0020}_{-  0.0017}$ &
$  0.0425^{+  0.0035}_{-  0.0027}$ &CFHTLenS + WMAP7 + R11 + BOSS
\\ \hline
$n_s$&
$  0.967^{+  0.013}_{-  0.013}$ &
$  0.965^{+  0.014}_{-  0.014}$ &
$  0.967^{+  0.014}_{-  0.014}$ &
$  0.967^{+  0.014}_{-  0.013}$ &CFHTLenS + WMAP7
\\ 
&
$  0.971^{+  0.011}_{-  0.012}$ &
$  0.964^{+  0.013}_{-  0.014}$ &
$  0.968^{+  0.014}_{-  0.014}$ &
$  0.966^{+  0.013}_{-  0.015}$ &CFHTLenS + WMAP7 + R11
\\ 
&
$  0.961^{+  0.012}_{-  0.011}$ &
$  0.957^{+  0.014}_{-  0.013}$ &
$  0.968^{+  0.013}_{-  0.014}$ &
$  0.961^{+  0.015}_{-  0.013}$ &CFHTLenS + WMAP7 + R11 + BOSS
\\ \hline
$\tau$&
$  0.089^{+  0.015}_{-  0.014}$ &
$  0.089^{+  0.016}_{-  0.014}$ &
$  0.088^{+  0.018}_{-  0.014}$ &
$  0.089^{+  0.016}_{-  0.014}$ &CFHTLenS + WMAP7
\\ 
&
$  0.092^{+  0.015}_{-  0.014}$ &
$  0.089^{+  0.016}_{-  0.014}$ &
$  0.089^{+  0.018}_{-  0.014}$ &
$  0.088^{+  0.016}_{-  0.013}$ &CFHTLenS + WMAP7 + R11
\\ 
&
$  0.082^{+  0.014}_{-  0.012}$ &
$  0.082^{+  0.017}_{-  0.012}$ &
$  0.086^{+  0.016}_{-  0.011}$ &
$  0.084^{+  0.016}_{-  0.013}$ &CFHTLenS + WMAP7 + R11 + BOSS
\\ \hline
$\Delta^2_R$&
$  2.395^{+  0.086}_{-  0.087}$ &
$  2.405^{+  0.094}_{-  0.086}$ &
$  2.412^{+  0.090}_{-  0.096}$ &
$  2.430^{+  0.103}_{-  0.096}$ &CFHTLenS + WMAP7
\\ 
&
$  2.378^{+  0.079}_{-  0.086}$ &
$  2.412^{+  0.098}_{-  0.082}$ &
$  2.418^{+  0.090}_{-  0.098}$ &
$  2.420^{+  0.095}_{-  0.090}$ &CFHTLenS + WMAP7 + R11
\\ 
&
$  2.427^{+  0.092}_{-  0.072}$ &
$  2.440^{+  0.083}_{-  0.093}$ &
$  2.382^{+  0.102}_{-  0.091}$ &
$  2.391^{+  0.111}_{-  0.072}$ &CFHTLenS + WMAP7 + R11 + BOSS
\\ \hline
\end{tabular}

\label{tab:cospar}
\end{table*}

The figures that follow in this section all compare constraints for different combinations of cosmological parameters and cosmological models with the following colour-scheme: WMAP7-only (in blue), WMAP7 combined with CFHTLenS and R11 (in pink),  WMAP7 combined with BOSS and R11 (in green) and all four data sets in combination (in white).  Comparing the green contours with the pink contours allows the reader to gauge the relative power of BOSS and CFHTLenS when either survey is used in combination with WMAP7 and R11.  Comparing the green contours with the white contours allows the reader to gauge the extra contribution that CFHTLenS makes to BOSS, R11 and WMAP7 in breaking different parameter degeneracies and constraining cosmological parameters.  

\subsubsection{Constraints in the $\sigma_8-\Omega_{\rm m}$ plane}
\label{sec:s8omcon}
\begin{figure*}
  \centering
   \includegraphics[width=7.0in, angle=0]{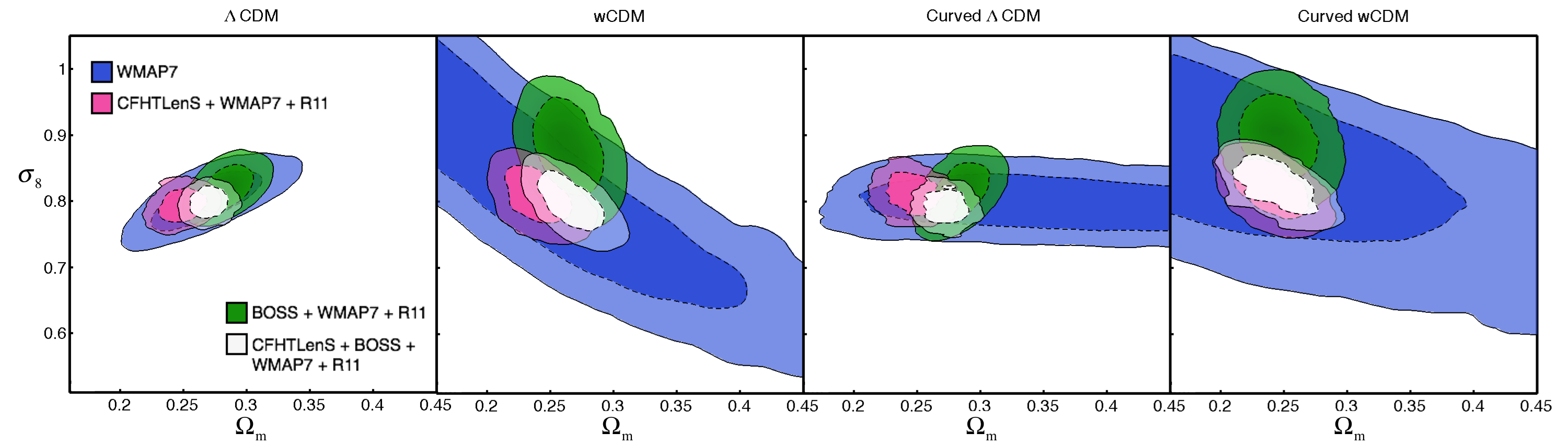} 
   \caption{Joint parameter constraints on the normalization of the matter power spectrum $\sigma_8$ and the matter density parameter $\Omega_{\rm m}$ from WMAP7-only (blue), BOSS combined with WMAP7 and R11 (green), CFHTLenS combined with WMAP7 and R11 (pink) and CFHTLenS combined with BOSS, WMAP7 and R11 (white).}   
   \label{fig:s8om_allcosmo}
\end{figure*}

Figure~\ref{fig:s8om_allcosmo} shows joint parameter constraints on the normalisation of the matter power spectrum $\sigma_8$ and the matter density parameter $\Omega_{\rm m}$ for four cosmological models:  flat $\Lambda$CDM, flat $w$CDM, curved $\Lambda$CDM and curved $w$CDM.   The comparison of the results for the four cosmological models show the decreased WMAP7 sensitivity to these two cosmological parameters when extra freedom in the cosmological model is introduced, such as dark energy $w_0$, or curvature.   We find slightly tighter constraints from CFHTLenS in combination with WMAP7 and R11 (pink), in comparison to BOSS in combination with WMAP7 and R11 (green).   The 68 per cent confidence regions between these two survey combinations only marginally overlap, introducing a mild tension.  The constraints are however consistent at the 95 per cent confidence level.  For the matter density parameter $\Omega_{\rm m}$, the addition of BOSS data to the combined CFHTLenS, WMAP7, R11 analysis typically decreases the 1$\sigma$ errors by $\sim 20$ per cent across all cosmologies.   For the normalisation of the matter power spectrum $\sigma_8$, however, we find BOSS adds little to the constraining power of CFHTLenS with WMAP7 and R11 for the cosmological models tested.  Furthermore, for a flat $\Lambda$CDM cosmology the constraint $\sigma_8 = 0.799 \pm 0.015$ is 
almost entirely driven by CFHTLenS in combination with WMAP7 alone.    

\subsubsection{Curved cosmological models}
\begin{figure*}
   \centering
   \includegraphics[width=6.5in, angle=0]{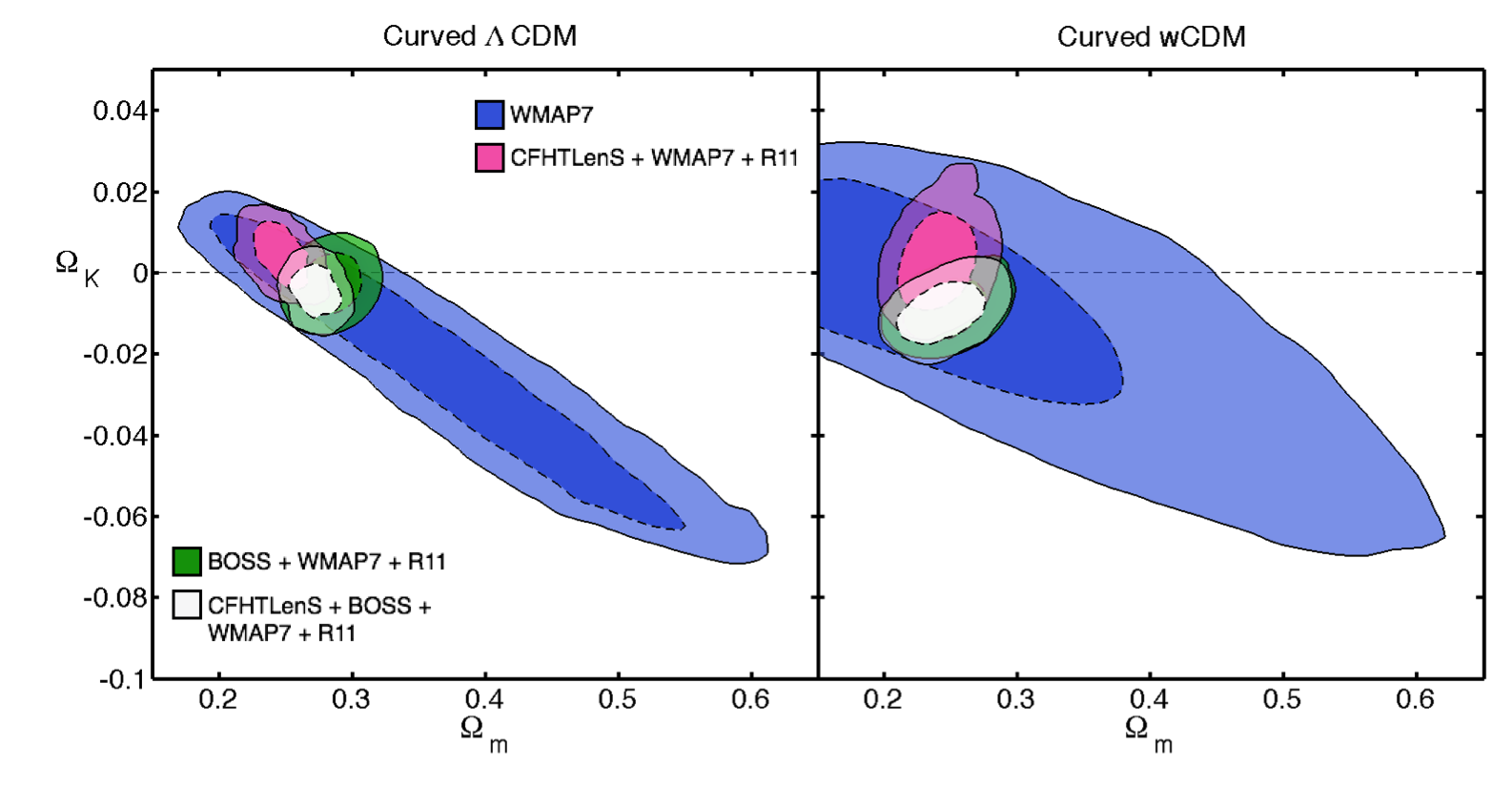} 
   \caption{Joint parameter constraints on curvature showing constraints on the curvature parameter $\Omega_{\rm K}$ and the matter density parameter $\Omega_{\rm m}$ from WMAP7-only (blue), BOSS combined with WMAP7 and R11 (green), CFHTLenS combined with WMAP7 and R11 (pink) and CFHTLenS combined with BOSS, WMAP7 and R11 (white).}   
   \label{fig:odeom}
\end{figure*}

We consider two curved cosmologies where the sum of the different density components of the Universe is no longer limited to the critical density.  Figure~\ref{fig:odeom} shows joint parameter constraints on the curvature $\Omega_{\rm K}$ and the matter density parameter $\Omega_{\rm m}$ for WMAP7-only (blue), BOSS combined with WMAP7 and R11 (green), CFHTLenS combined with WMAP7 and R11 (pink) and CFHTLenS combined with BOSS, WMAP7 and R11 (white).  In both the curved $\Lambda$CDM and curved $w$CDM cosmology we find that the data are consistent with a flat Universe with $\Omega_K \simeq -0.004 \pm 0.004$ (see Table~\ref{tab:cospar} for exact numbers for the different cosmologies and data combinations).   In this parameter space we find a factor of two improvement when R11 is included in combination with CFHTLenS and WMAP7.  This is partly because when curvature is allowed the degeneracy direction of the CMB in the $\sigma_8-\Omega_{\rm m}$ plane changes such that the combination of lensing with the CMB becomes less powerful.  Little improvement is found in the constraining power when BOSS is included in our parameter combination, but the mean $\Omega_K$ changes by nearly $2\sigma$.

\subsubsection{Constraints on dark energy}
\begin{figure*}
   \centering
   \includegraphics[width=7.0in, angle=0]{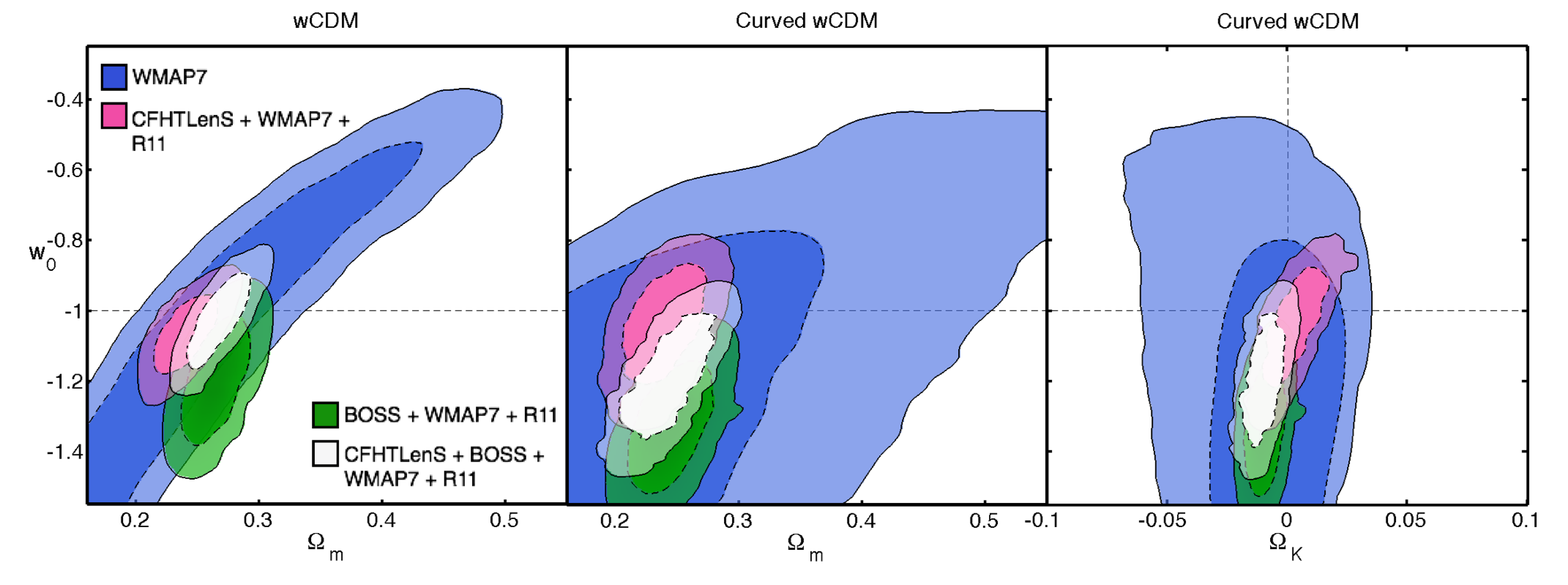} 
   \caption{Joint parameter constraints on  the dark energy equation of state parameter $w_0$ and the matter density parameter $\Omega_{\rm m}$, and  curvature parameter $\Omega_{\rm K}$ for a curved $w$CDM cosmology from WMAP7-only (blue), BOSS combined with WMAP7 and R11 (green), CFHTLenS combined with WMAP7 and R11 (pink) and CFHTLenS combined with BOSS, WMAP7 and R11 (white).}   
   \label{fig:wodeom}
\end{figure*}

Finally we turn to the constraints that can be placed on the dark energy equation of state parameter $w_0$ in flat and curved cosmologies.  Figure~\ref{fig:wodeom} shows joint parameter constraints in the $w - \Omega_{\rm m}$ plane and also the $w - \Omega_{\rm K}$  plane for a curved $w$CDM cosmology.  As with the other parameter planes that we have commented upon in this Section, we again see the mild tension between BOSS and CFHTLenS and the power of including these surveys in addition to WMAP7 data alone.  For both the curved and flat $w$CDM cosmologies we find that $w$ is consistent with a cosmological constant (see Table~\ref{tab:cospar} for exact numbers for the different cosmologies and data combinations).    As with our constraints on the curvature, we find very little improvement in the constraining power on $w$ when BOSS is included in our parameter combination.  We do, however, find excellent agreement with the combined probe constraints from \citet{BOSS} when BOSS is combined with WMAP7, SDSS-LRG baryon acoustic oscillation constraints from \citet{Pad} and Type Ia supernovae results from \citet{SNLS3}.  With this combination of data sets,  \citet{BOSS} find $w = -1.09 \pm 0.08$ for flat and curved $w$CDM models.  Note that this was shown to be the only parameter where the addition of the supernova data to the BOSS and WMAP7 data impacted upon the analysis, decreasing the errors by a factor of $\sim 2$.  This result is in agreement with our $w$CDM model constraints from CFHTLenS with WMAP7 and R11, where we find  $w = -1.06 \pm 0.08$ (flat) and $w = -1.04 \pm 0.12$ (curved). 

We find good agreement between the mean measurements when the different parameters sets are combined.  This is in contrast to the 2D weak lensing analysis of \citet{Kilbinger2012} where a $2\sigma$ difference is found between the mean $w_0$ measured with lensing, WMAP7 and BOSS, with and without a prior on the hubble parameter $h$.  For all cosmologies tested in this analysis, the constraints on $h$ from CFHTLenS with WMAP7 are in good agreement with the R11 measure of $h = 0.738 \pm 0.024$.  Focussing on flat $\Lambda$CDM, in this analysis we find $h=0.717 \pm 0.016$ for CFHTLenS with WMAP7, in comparison to BOSS with WMAP7 who find a $2\sigma$ offset from R11 with $h=0.684 \pm 0.013$.  For $w$CDM cosmologies,  \citet{Kilbinger2012} and BOSS find even larger shifts away from the R11 result, but at a lower significance, and it is this that causes the differences in the measurement of $w_0$, with and without the inclusion of a prior on $h$, that we do not find in this analysis.

\section{The Intrinsic Alignment of early-type and late-type galaxies}
\label{sec:IAredblue}

\begin{figure}
   \centering
   \includegraphics[width=3.0in, angle=270]{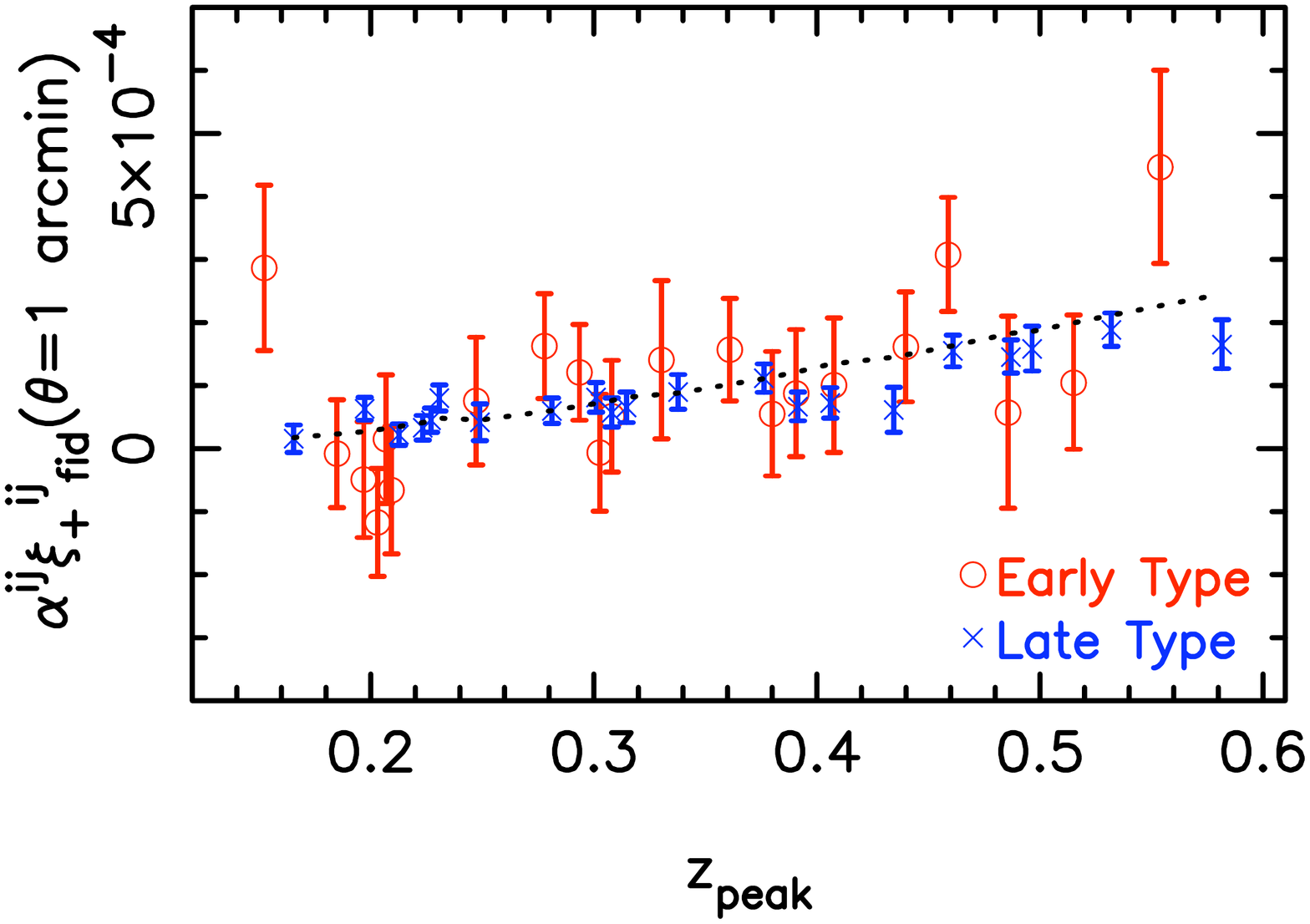} 
      \caption{Compressed CFHTLenS tomographic data for two galaxy samples; early-type (circles) and late-type (cross) galaxies.  As in Figure~\ref{fig:datacomp}, each point represents a different tomographic bin combination $ij$ as indicated by $z_{\rm peak}$, the peak redshift of the lensing efficiency for that bin.  The measured best-fitting amplitude $\alpha^{ij}$ of the data for each galaxy type, multiplied by the fiducial model at $\theta = 1$ arcmin for $\xi_+$. is shown.  The error bars show the $1\sigma$ constraints on the fit.  The data can be compared to the fiducial GG-only model, shown dotted.  }   
   \label{fig:redblue}
\end{figure}

\begin{figure}
   \centering
   \includegraphics[width=3.0in, angle=270]{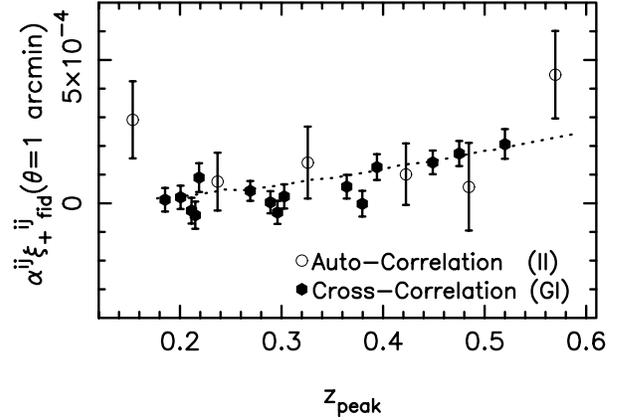} 
      \caption{Compressed CFHTLenS tomographic data for an optimised early-type galaxy intrinsic alignment measurement with auto-correlated redshift bins containing only early-type galaxies (circles) and cross-correlation redshift bins containing early-type galaxies in the low redshift bin and all galaxy types in the high redshift bin (filled).  Different tomographic bin combinations $ij$ are indicated by $z_{\rm peak}$, the peak redshift of the lensing efficiency for that bin.  The best-fitting amplitude $\alpha^{ij}$ of the data relative to a fixed fiducial GG-only cosmology model is shown, multiplied by the fiducial model at $\theta = 1$ arcmin for $\xi_+$.  The error bars show the $1\sigma$ constraints on the fit.  The data can be compared to the fiducial GG-only model, shown dotted.  }   
   \label{fig:redopt}
\end{figure}

As discussed in Section~\ref{sec:intro} there is clear evidence in the literature that the strength of the intrinsic alignment signal depends on galaxy type, with the most massive red galaxies exhibiting the strongest intrinsic correlations \citep{BJ11}.   In this section we therefore present separate tomographic analyses of an early-type and late-type galaxy sample, selected using the measured best-fit photometric type $T_{\rm BPZ}$.  This classification type ranges from 1 to 6 and represents the best-fitting spectral energy distribution to each galaxy's photometry \citep[see][for more details]{HH12}.  We follow \citet{Simon2012} by selecting late-type spiral galaxies with $T_{\rm BPZ}>2.0$, roughly 80 per cent of the galaxy catalogue used in the main analysis.  The remaining 20 per cent are classified as early-type galaxies.  Each sample is split into 6 tomographic bins, using the redshift selection given in Table~\ref{tab:zbins}, and the redshift distribution determined from the sum of the $P(z)$.  Covariance matrices were determined for each galaxy sample using the method outlined in Section~\ref{sec:covest}, but mapping only the relevant galaxy sample on to the N-body lensing simulations.  We found no evidence for a significant difference in $\sigma_e$ for the two samples.  Figure~\ref{fig:redblue} shows the resulting compressed tomographic measurements made with early-type galaxies (circles) and late-type galaxies (crosses).  The data compression  uses the visualization method described in Section~\ref{sec:vis}, modified slightly such that the free amplitude parameter $\alpha^{ij}$ is fitted simultaneously to both the $\xi_+$ and $\xi_-$ measured from the data.   This simultaneous fit is justified as the $\alpha^{ij}$ values measured for $\xi_+$ and $\xi_-$ independently are fully consistent.  The resulting best-fitting amplitude $\alpha^{ij}$ is shown, multiplied by the fiducial model at $\theta = 1$ arcmin for $\xi_+$.
With only 20 per cent of the data contained in the early-type sample, it is unsurprising that the measured signal to noise is significantly weaker than for the late-type sample which are well fit by the fiducial GG-only model, shown dotted.    We can, however, optimise the measurement of the intrinsic alignment signal from early-type galaxies, to get a clearer picture, if we assume the II contribution to cross-correlated bins is small in comparison to the GI signal.  If this is the case, we can decrease the noise on the GI measurement by using the full galaxy sample as background galaxies to correlate with the early-type galaxies in the foreground bin.  The result of this optimised analysis is shown, in compressed tomographic data form, in Figure~\ref{fig:redopt}.  The open circles show the tomographic signal measured in the auto-correlated redshift bins between early-type galaxies (these auto-correlation bins are also shown in Figure~\ref{fig:redblue}).   The closed symbols show the tomographic signal in the cross-correlated redshift bins where early-type galaxies populate the foreground bin and the full galaxy sample populates the background higher redshift bin.  The data can be compared to the fiducial GG-only model, shown dotted.  What is interesting to note from this Figure is that at low redshifts, where the intrinsic alignment signal is expected to be the most prominent, the auto-correlated bins tend to lie above the GG-only model.  We expect this from the II term.  For the cross-correlated bins, however, the measured signal tends to lie below the GG-only model.  We expect this from the GI term.

\begin{figure*}
   \centering
   \includegraphics[width=6.8in, angle=0]{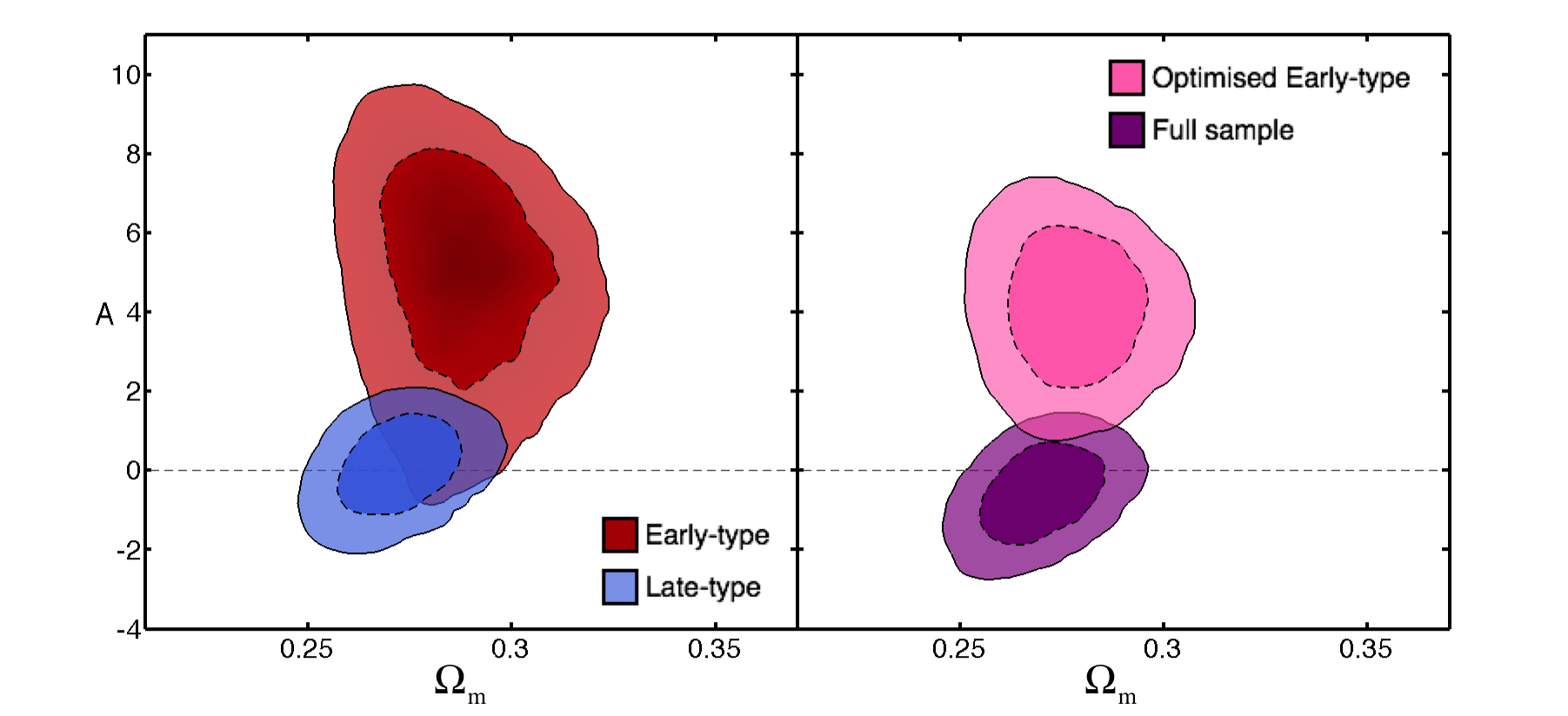} 
   \caption{Joint parameter constraints on the amplitude of the intrinsic alignment model $A$ and the matter density parameter $\Omega_{\rm m}$ from CFHTLenS combined with WMAP7, BOSS and R11.  In the left panel the constraints can be compared between two galaxy samples split by SED type, (early-type in red and late-type in blue).  In the right panel we present constraints from a optimised analysis to enhance the measurement of the intrinsic alignment amplitude of early-type galaxies (pink).  The full sample, combining early and late-type galaxies, produces an intrinsic alignment signal that is consistent with zero (shown purple).  A flat $\Lambda$CDM cosmology is assumed.}   
   \label{fig:Aomegaredblue}
\end{figure*}

Figure~\ref{fig:Aomegaredblue} combines the CFHTLenS data split by galaxy type, and our optimised early-type galaxy tomography analysis, with auxiliary data from WMAP7, BOSS and R11 to constrain the amplitude of the intrinsic alignment model $A$.  Assuming a flat $\Lambda$CDM model, the resulting 68 per cent and 95 per cent confidence limits on $A$ and the matter density parameter $\Omega_{\rm m}$ can be compared\footnote{Note that the constraints on cosmological parameters other than $A$ are consistent between the early-type and late-type analysis, and that both sets of parameter constraints, with the exception of $A$, are consistent with the full galaxy sample analysis reported in table~\ref{tab:cospar}.}.  In the left panel we show constraints from the two galaxy samples split by SED type.  The early-type galaxy constraints are shown in red and the late-type galaxy constraints are shown in blue.   In the right panel, constraints are shown for the full galaxy sample in purple and  the optimised early-type intrinsic alignment analysis in pink.
The marginalised 68 per cent confidence errors on $A$, from the combination of CFHTLenS data with WMAP7, BOSS and R11, for the four different measurements are
\be
A_{\rm late} = 0.18^{+ 0.83}_ {- 0.82}\, ,
\ee
\be
A_{\rm early} = 5.15^{+ 1.74}_{- 2.32}\, ,
\ee
\be
A_{\rm early}^{\rm opt} = 4.26^{+ 1.23}_{- 1.39}\, ,
\ee
\be
A_{\rm all}  = -0.48^{+ 0.75}_{- 0.87}  \, . 
\ee
We find the intrinsic alignment amplitude of the late-type sample is consistent with zero.  In contrast, the amplitude of the intrinsic alignment model for the early-type sample is detected to be non-zero with close to $2\sigma$ confidence.  When we consider the optimised analysis, we find an even stronger detection, with an intrinsic alignment amplitude of $A=0$ for early-type galaxies ruled out with $3\sigma$ confidence.  The optimised early-type analysis should be considered with some caution, however, as the tomographic redshift bins do overlap and as such a small fraction of late-type with early-type II correlation will be included in the measurement.  The measurement of $A_{\rm early}$ should therefore be considered as our cleanest measurement of the early-type galaxy intrinsic amplitude with the optimised  $A_{\rm early}^{\rm opt}$ analysis providing us with the strongest evidence for intrinsic galaxy alignments between early-type galaxies.

\subsection{Discussion}
Our constraints show the same broad findings as other studies; intrinsic alignments are dependent on galaxy type.  As previous studies have focused on specific galaxy samples at fixed redshifts, however, it is difficult to compare our constraints directly.  With that caveat we can, however, comment on literature results from galaxy samples that are the most comparable.  Our late-type sample is most similar in its properties to the blue galaxies from the WiggleZ survey analysed in \citet{RM11}.  Their null detection is in agreement with our late-type galaxy results.  Our early-type sample is most similar in terms of luminosity and redshift to the MegaZ-LRG sample analysed in \citet{BJ11}. The best-fit values $4 \ls A \ls 6$ for a range of different types of LRG galaxy selection with an error of $\sim 1$, are in very good agreement with our early-type galaxy results.  

For the full galaxy sample, there is an indication that negative values of $A$ are preferred.  For flat cosmologies, $A$ is negative at the $1.4\sigma$ level when the CFHTLenS data are combined only with WMAP7 and R11 (see table~\ref{tab:cospar} for constraints on $A$ for the full galaxy sample for different cosmologies and data combinations).  Whilst we emphasize that this result is not statistically significant it is however worth commenting on what this finding could mean.  In the conventional intrinsic alignment model the GI signal is negative and scales with $A$.  The II signal is positive and scales with $A^2$.  Finding $A<0$, however, implies the data prefer a GI+II signal that is more positive than the conventional model would predict.  This suggests that future surveys with lower statistical errors, should aim to fit independent amplitudes to the GI and II signals as the interplay between the two effects may be more complex than the linear tidal field alignment model suggests.

It is also interesting to comment on the decrease in the amplitude of the best-fit intrinsic alignment signal when early and late type galaxies are combined.  If detected in future surveys at higher significance, this would indicate a complex interplay between the two galaxy types.  It has long been thought that the reason for the difference between the intrinsic alignments of early and late type galaxies lies in the different mechanisms at play during galaxy formation.  The intrinsic alignment model we use in this analysis is based on linear theory.  A more traditional galaxy formation scenario for late-type galaxies, however,  is tidal-torque theory where it is the angular momentum of the dark matter field that induces galaxy spin and hence intrinsic galaxy alignments \citep[see][and references therein]{Schafer09}.  The simple hypothesis, presented in \citet{HeymansIA06}, is that the intrinsic alignment of early-type galaxies is a result of ellipticity deformations due to the linear tidal field, in contrast to late-type galaxies whose alignment results from angular momentum induced ellipticity alignments \citep{vdBosch02}.  This hypothesis is in good agreement with recent observations of galaxy-type dependence in the intrinsic alignment signal, as halo angular momentum is proportional to the square of the tidal shear, and the induced galaxy alignments therefore correlate over much shorter ranges compared to alignments directly caused by the linear tidal shear \citep{CKB01}.  

In addition to the linear model used throughout this paper, \citet{HS04} also investigate the GI signal expected from an intrinsic alignment model where the galaxy ellipticity is proportional to the square of the tidal field.  In this case the GI signal is expected to be zero.  As our galaxy sample is dominated by late-type galaxies, the majority of correlated galaxy pairs in our analysis from different redshift bins will include a late-type foreground galaxy.  Combining the findings of \citet{HS04} with our simple hypothesis that late-type intrinsic galaxy alignment is caused by halo angular momentum induced alignments, leads to an expected zero GI measurement on average.  In auto-correlated tomographic bins however, the stronger galaxy clustering of early-type galaxies will mean that at small angular scales, there is a higher proportion of close early-type galaxy pairs in the measurement, compared to the numbers of early-type and late-type foreground galaxies that contribute to the GI signal.  This therefore boosts the true II signal in auto-correlated bins over the amplitude that would be predicted from GI-only constraints from a mixed galaxy population.

The linear tidal field alignment model used in this analysis could compensate for these different galaxy-type contributions to the II and GI signal by favoring a small but negative value for $A$.  In this case the GI signal in the cross-correlation bins is positive but sufficiently weak to provide a reasonable fit to the GI=0 model signal expected from the dominant late-type galaxy population.  In the auto-correlated bins, the additional true positive II signal from the clustered early-type galaxies is then represented in the model fit, not by the model II signal, but the positive GI signal.  If $A$ was positive and less than unity, there would still be a reasonable weak but now negative fit to the GI=0 model in the cross-correlation bins.  In the auto-correlated bins, however, there would not be sufficient signal in the combined II+GI model to represent the extra II power arising from the clustered early-type sample.  

Based on this discussion, we can conclude that our constraints for the full sample favoring a slightly negative value for $A$ fits our simple hypothesis that early-type galaxy alignment results from the linear tidal field and late-type galaxy alignment results from angular momentum induced correlations.  The next generation of weak lensing surveys will have the statistical power to test this hypothesis further.

\section{Conclusions}
\label{sec:conc}
The Canada-France-Hawaii Telescope Lensing Survey, CFHTLenS, represents the current state-of-the-art in cosmological weak lensing data analysis, from the applied weak lensing optimised data reduction, shear and photometric redshift measurement methods, through to the robust systematic error analysis and error quantification of the resulting shear and redshift catalogue.  Spanning 154 square degrees, CFHTLenS is currently the largest deep weak lensing survey in existence permitting the tightest cosmological constraints from weak gravitational lensing.  In this article we present the first multi-redshift bin, or tomographic, weak lensing analysis to mitigate the contamination to the measured two-point shear correlation function through the simultaneous fit of a cosmological model with an intrinsic galaxy alignment model.  Combining the tomographic CFHTLenS data with auxiliary cosmological probes; the cosmic microwave background with data from WMAP7, baryon acoustic oscillations with data from BOSS, and a prior on the Hubble constant from the HST distance ladder, we have improved constraints on a range of cosmological parameters for a standard flat $\Lambda$CDM model, in addition to curved and dark energy models.  We constrain the amplitude of the matter power spectrum $\sigma_8 = 0.799 \pm 0.015$ and the matter density parameter $\Omega_{\rm m} = 0.271 \pm 0.010$ for a flat $\Lambda$CDM cosmology.  For a flat $w$CDM cosmology we constrain the dark energy equation of state parameter $w = -1.02 \pm 0.09$.  In general we find tighter constraints from the combination of CFHTLenS with WMAP7 and R11 than from BOSS with WMAP7 and R11 and we find that the addition of BOSS to CFHTLenS with WMAP7 and R11 only significantly improves constraints on the matter density parameter $\Omega_{\rm m}$, for all cosmologies tested.  Constraints on the other parameters are only shown to significantly improve when a curved $w$CDM model is considered.    Finding consistent results, however, between these two very different probes of cosmology suggests a bright future for studies of the `Dark Universe' with weak lensing and baryon acoustic oscillations.

Tomographic weak lensing has long been recognized as a powerful tool to constrain dark energy by detecting the influence dark energy has on the growth of structure in addition to the distance-redshift relationship.  One astrophysical source of uncertainty that mimics cosmological weak lensing is the intrinsic alignment of neighbouring galaxies.  This phenomenon unfortunately reduces the overall constraining power of tomographic weak lensing analyses as, to ensure the cosmological constraints are unbiased, the contamination from intrinsic alignments must be considered.  In this analysis we have assumed a simple one-parameter model that scales the amplitude of the II and GI intrinsic alignment contamination expected from a linear tidal field alignment model of galaxy shape correlation \citep{CKB01, HS04, BK07}.  Our results are consistent with there being zero intrinsic alignment between galaxies in our late-type sample, with $A \simeq 0.2 \pm 0.8$.  The best-fit amplitude is, however, within $1\sigma$ of the fiducial model amplitude $A=1$.  This fiducial model is commonly assumed for parameter forecasts, and is based on the amplitude of low-redshift galaxy intrinsic ellipticity correlations measured by \citet{BTHD02}.  For the 20 per cent of galaxies in our sample whose 5-band photometry is best-fit by an early-type SED, however, we detect a non-zero intrinsic alignment signal, $ A \simeq 5 \pm 2$, roughly five times the fiducial model amplitude.  This is in agreement with previous observations of a galaxy-type dependence, using a very different methodology \citep{Hirata07, BJ11,RM11}.  Our results therefore add to the increasing body of independent observations that point towards a scenario where the galaxy formation and evolution mechanisms, which determine galaxy shape, differ for early and late-type galaxies.  For the combined galaxy sample, we find that the net effect of the two galaxy types produces an intrinsic alignment signal that is consistent with zero, with $A \simeq-0.5 \pm 0.8$, (see Section~\ref{sec:res} for the exact constraints for the different cosmological models).

One difficulty, for the analysis of finely-binned tomographic correlation functions which are optimised for the simultaneous analysis of cosmological models and intrinsic alignment models, is the estimation of an accurate and invertible covariance matrix for the large data vector.  In this paper we discuss the different options for covariance matrix estimation and the noise biases that can arise in the matrix inversion.  We chose to use N-body lensing simulations as the basis for our covariance matrix estimation in order to correctly account for non-Gaussianity on small angular scales.  This is in contrast to many earlier studies which often resorted to a fitting function correction.  This choice, however, limits the largest angular scales that we can analyse.  The finite simulated box size truncates the large-scale modes, reducing the large scale power probed by each simulated line of sight.  In addition, as high-resolution N-body lensing simulations are expensive to create, we are also limited by the total number of simulated independent lines of sight available.  This finite number limits the number of data points for which we can invert the simulation estimated covariance matrix without biasing our results, or erroneously increasing the area of the resulting confidence regions.  A priority of future surveys must therefore be to ensure the availability of a large volume of N-body simulations for covariance matrix estimation.  Alternative methods should also be developed such as the simulation-analytic covariance matrix grafting method described in \citet{Kilbinger2012}, in addition to methods which optimally combine simulation and data-based bootstrap or jackknife estimates.

Three major new weak lensing surveys\footnote{KiDS: kids.strw.leidenuniv.nl, DES:  www.darkenergysurvey.org and HSC: www.subarutelescope.org/Projects/HSC} are currently in progress, or will commence soon, and as such this is an exciting time for the study of the `Dark Universe'.   These surveys will image more than ten times the area surveyed by CFHTLenS, charting a sufficient volume to address the issue of astrophysical bias arising from intrinsic galaxy alignments in greater detail.  By self-calibrating the intrinsic alignment model with the additional information from galaxy clustering \citep{Bernstein09, Zhang10,JBSB10}, the marginalisation over many intrinsic alignment nuisance parameters will become feasible. Another viable mitigation strategy is to develop new statistics which are less sensitive to intrinsic alignment contamination.   In addition, complementary spectroscopic or highly accurate photometric redshift observations could be exploited to place observational constraints on different analytical and hydro-dynamical simulation models of galaxy shape correlations.  This effort, in parallel to large-area survey acquisition, will then allow us retain as much power as possible from weak gravitational lensing as a cosmological probe.

\section{Acknowledgements}
We would like to thank Benjamin Joachimi for the verification of our non-linear intrinsic alignment model theory code in addition to many constructive comments during the preparation of this paper.  We also thank the referee, in addition to Sarah Bridle and Donnacha Kirk for helpful discussions about intrinsic alignment modeling.  The visualization of the Monte-Carlo samples and cosmological parameter constraints makes use of code adapted from CosmoloGUI\footnote{CosmoloGUI: www.sarahbridle.net/cosmologui}.

This work is based on observations obtained with MegaPrime/MegaCam, a joint project of CFHT and CEA/DAPNIA, at the Canada-France-Hawaii Telescope (CFHT) which is operated by the National Research Council (NRC) of Canada, the Institut National des Sciences de l'Univers of the Centre National de la Recherche Scientifique (CNRS) of France, and the University of Hawaii. This research used the facilities of the Canadian Astronomy Data Centre operated by the National Research Council of Canada with the support of the Canadian Space Agency.  We thank the CFHT staff for successfully conducting the CFHTLS observations and in particular Jean-Charles Cuillandre and Eugene Magnier for the continuous improvement of the instrument calibration and the {\sc Elixir} detrended data that we used. We also thank TERAPIX for the quality assessment and validation of individual exposures during the CFHTLS data acquisition period, and Emmanuel Bertin for developing some of the software used in this study. CFHTLenS data processing was made possible thanks to significant computing support from the NSERC Research Tools and Instruments grant program, and to HPC specialist Ovidiu Toader.  The N-body simulations used in this analysis were performed on the TCS supercomputer at the SciNet HPC Consortium. SciNet is funded by: the Canada Foundation for Innovation under the auspices of Compute Canada; the Government of Ontario; Ontario Research Fund - Research Excellence; and the University of Toronto.  The early stages of the CFHTLenS project were made possible thanks to the support of the European CommissionÕs Marie Curie Research Training Network DUEL (MRTN-CT-2006-036133) which directly supported members of the CFHTLenS team (LF, HHi, BR, MV) between 2007 and 2011 in addition to providing travel support and expenses for team meetings.

CH, EG, FS, HHo, ES, MLB \& BR acknowledge support from the European Research Council under the EC FP7 grant numbers 240185 (CH, EG, FS), 279396 (HHo, ES), 280127 (MLB) \& 240672 (BR).  EG also acknowledges the award of an STFC studentship.   TDK acknowledges support from a Royal Society University Research Fellowship.  TE is supported by the Deutsche Forschungsgemeinschaft through project ER 327/3-1 and is supported by the Transregional Collaborative Research Centre TR 33 - "The Dark Universe". HHi is supported by the Marie Curie IOF 252760, a CITA National Fellowship, and the DFG grant Hi 1495/2-1. HHo also acknowledges support from Marie Curie IRG grant 230924 and, with ES, the Netherlands Organisation for ScientiÞc Research grant number 639.042.814.    YM acknowledges support from CNRS/INSU (Institut National des Sciences de l'Univers) and the Programme National Galaxies et Cosmologie (PNCG).  LVW acknowledges support from the Natural Sciences and Engineering Research Council of Canada (NSERC) and the Canadian Institute for Advanced Research (CIfAR, Cosmology and Gravity program).  MLB also acknowledges the award of an STFC Advanced Fellowship (ST/I005129/1). 
LF acknowledges support from NSFC grants 11103012 \& 10878003, Innovation Program 12ZZ134, Chen Guang project 10CG46 of SMEC, STCSM grant 11290706600 and the Pujiang Program 12PJ1406700.   MJH acknowledges support from the Natural Sciences and Engineering Research Council of Canada (NSERC).  TS acknowledges support from NSF through grant AST-0444059-001, SAO through grant GO0-11147A, and NWO. MV acknowledges support from the Netherlands Organization for Scientific Research (NWO) and from the Beecroft Institute for Particle Astrophysics and Cosmology.  

{\small Author Contributions: All authors contributed to the development and writing of this paper.  The authorship list reflects the lead authors of this paper (CH, EG, AH, MK, TK, FS) followed by two alphabetical groups.  The first alphabetical group includes key contributers to the science analysis and interpretation in this paper, the founding core team and those whose long-term significant effort produced the final CFHTLenS data product.  The second group covers members of the CFHTLenS team who made a significant contribution to the project, this paper, or both.  CH and LVW co-led the CFHTLenS collaboration and TK led the CFHTLenS cosmology working group.}

\bibliographystyle{mn2e}
\bibliography{ceh_2012}
\label{lastpage}

\end{document}